\documentclass[final,times,twocolumn]{elsarticle}

\usepackage[utf8]{inputenc}
\usepackage{tabularx}
\usepackage{subfigure}
\usepackage{multirow}
\usepackage{fontawesome}
\usepackage{enumitem,array}
\usepackage{comment}
\usepackage{color}

\usepackage{amsmath}
\usepackage{xurl}
\usepackage{amssymb}
\usepackage{listings}

\usepackage{listings}
\usepackage{xcolor}
\usepackage{blindtext}
\usepackage{hyperref}
\usepackage{lineno}

\newcommand{\answer}[2]{
    \vspace{0.5cm}
    \noindent
    \fbox{\begin{minipage}{0.98 \columnwidth}
    {\textbf{RQ#1}: \textit{#2}}
    \end{minipage}
    }
    \vspace{0.1cm}
}

\AtBeginDocument{%
  \providecommand\BibTeX{{%
    \normalfont B\kern-0.5em{\scshape i\kern-0.25em b}\kern-0.8em\TeX}}}

\journal{ArXiv}

\begin{document}

\begin{frontmatter}

\title{Feature-oriented Test Case Selection and Prioritization During the Evolution of Highly-Configurable Systems}

\author[ufpr,bpk]{Willian D. F. Mendonça\corref{mycorrespondingauthor}}
\cortext[mycorrespondingauthor]{Corresponding author}
\ead{willianmendonca@ufpr.br}

\author[ncsu]{Wesley K. G. Assun\c{c}\~ao}
\ead{wguezas@ncsu.edu}

\author[ufpr]{Silvia R. Vergilio}
\ead{silvia@inf.ufpr.br}

\address[ufpr]{Computer Science Department, Federal University of Paraná (UFPR), \\CP: 19081, CEP: 81.531-980, Curitiba, Brazil}

\address[ncsu]{Department of Computer Science, North Carolina State University (NCSU), \\Raleigh, USA}

\address[bpk]{ Faculdade Biopark Educação, \\Toledo, Brazil}

\begin{abstract}
{\small  Testing {\it Highly Configurable Systems (HCSs)} is a challenging task, especially in an evolution scenario where features are added, changed, or removed, which hampers test case selection and prioritization. Existing work is usually based on the variability model, which is not always available or updated. Yet, the few existing approaches rely on links between test cases and changed files (or lines of code), not considering how features are implemented, usually spread over several and unchanged files. To overcome these limitations, we introduce  \texttt{FeaTestSelPrio}, a feature-oriented test case selection and prioritization approach for HCSs.  The approach links test cases to feature implementations, using HCS pre-processor directives, to select test cases based on features affected by changes in each commit. After, the test cases are prioritized  according to the number of features they cover. Our approach selects a greater number of tests and takes longer to execute than  a changed-file-oriented approach, used as baseline, but \texttt{FeaTestSelPrio} performs better regarding detected failures. By adding the approach execution time to the execution time of the selected test cases, we reached a reduction of  $\approx$50\%, in comparison with retest-all.  The prioritization step allows reducing the average test budget in 86\% of the failed commits.}
\end{abstract}


\end{frontmatter}

\section{Introduction}
\label{section:Introduction}

\textit{Highly Configurable Systems (HCSs)} are pieces of software designed to promote customization and flexibility while leveraging systematic reuse~\cite{MichelonGPCE2021}. The systematic reuse is achieved by creating a common core of assets that is shared among different products that can be derived from an HCS~\cite{clements2002software}. The customization and flexibility comes from a set of configuration options, enabling different functionalities to be selected to a given context or needs~\cite{vonRhein2015}. For the design and implementation of HCSs, the configuration options are captured as \textit{features} of the target domain or market segment~\cite{Kang1990}. 

The main benefit of HCSs is to reduce the time to market of new products, by systematically reusing the core of assets~\cite{clements2002software}, as mentioned above. However, due to this extensive reuse, the testing of HCSs is a critical activity~\cite{do2014strategies}. A bug in a common feature (i.e., configuration option) can lead to failures in a potentially large number of software products~\cite{ferreira2019testing}. To make matters worse, HCSs usually have a substantial number of features and interrelated dependencies that increases the complexity of testing~\cite{Medeiros2018}.

Software testing for HCSs is even more challenging when we consider an evolution scenario, such as in the cases in which {\it Continuous Integration (CI)} practices are adopted~\cite{lima2020learning}. HCSs can be updated, integrated, and tested several times a day, and each cycle needs to be fast~\cite{zhao2017impact}.  CI environments automatically support tasks such as the build process, test execution, and test results reporting, allowing engineers to merge code that is under development or maintenance with the mainline code base at frequent time intervals \cite{duvall2007continuous}. The results are used to solve problems and find faults. 
Re-executing all test cases (i.e., \textit{retest-all} technique) during each evolution cycle of the HCS may be impracticable, as the test activity in an industrial environment is extremely costly~\cite{garousi2013survey}. Then, providing quick feedback is essential to reduce development costs~\cite{Jiang2016}. To this end, the application of {\it Regression Testing (RT)} techniques in a very cost-effective way is fundamental~\cite{garousi2013survey}. The best known and used RT techniques are \cite{Yoo2012}: (i) Test Case Minimization~(TCM) usually removes redundant test cases, minimizing the test set according to some criteria; (ii) Test Case Selection~(TCS) selects a subset of test cases, the most important for testing the software; and (iii) Test Case Prioritization~(TCP) attempts to reorder a set of tests to identify an ideal order that, ideally, maximizes early fault detection.

Existing TCS approaches for HCSs are usually based  on {\it Feature Model (FM)} or other artifacts representing variability~\cite{lity2019retest, lachmann2015delta, al2017delta, lachmann2017risk, neto2010regression}. However, none of them consider that HCSs are usually developed by adopting CI practices, in a scenario where the FMs are rarely  updated ~\cite{vierhauser2010flexible,ghanam2010linking,heider2012using}, making the use of those approaches difficult. Other HCS testing approaches rely on dynamic analysis based on the test failure-history, execution, or coverage~\cite{marijan2017titan, marijan2018practical, marijan2019learning}. In some pieces of work, the approaches are only evaluated with systems well-modularized~\cite{jung2019automated}, having test cases separated by feature and without overlap among features, which is different from real scenarios.
Testing approaches in CI are also based on code changes. For instance, TCS approaches select test cases related to the files changed in the current commit~\cite{gligoric2015practical, bertolino2020learning, romano2018spiritus}. But those existing approaches and tools do not consider HCS particularities (i.e., features as building blocks), and they are mostly based on Java language only~\cite{jung2019automated,jung2020efficient,jung2023automated}. 
However, C and C++ are the preferred language, largely adopted for  HCS developers~\cite{Medeiros2018, Michelon2022}.   Yet, the few existing approaches rely on links between test cases and files or lines of code, limiting the selection to test cases related to file changes, not considering the whole implementation of features, which can be spread over many files other than the changed ones.

Motivated by these facts, in a previous work, we introduced \texttt{FeaTestSel}~\cite{mendoncaTCS2023}, a feature-oriented approach for test case selection that links test cases to features using HCS pre-processor directives. Given the source code of an HCS and a set of test cases available for a given commit, \texttt{FeaTestSel} selects the best test cases to be executed in order to cover the features changed in the corresponding evolution cycle. \texttt{FeaTestSel} also produces different traceability reports linking, namely (i) test cases to code lines of the system, (ii) features to code lines of the system, and (iii) test cases to features.  Differently from related work, the implementation of \texttt{FeaTestSel} works for systems in the C/C++ language. Our approach is feature-oriented and needs only static analysis of the source code. 
Results from our previous work with the system \texttt{Libssh} show that \texttt{FeaTestSel} reduces the testing runtime to 
$\approx$50\% compared with the retest-all technique. Furthermore, the approach was able to maintain the test quality by selecting 100\% of test cases that revealed failures. The main advantage of a feature-oriented approach is to provide a natural way to think about the units to be tested, since features are the building blocks of HCSs, used as main units for HCS design and communication among stakeholders.

Leveraging contributions of our previous study, in the present work we propose an extension for  \texttt{FeaTestSel}, namely  \texttt{FeaTestSelPrio} ({\bf Fea}ture-oriented {\bf Test} Case {\bf Sel}ection and {\bf Pri}oritization for Highly Configurable Systems). This extension includes an additional step that performs the test case prioritization considering the reports generated by  \texttt{FeaTestSel}. Furthermore, new evaluation results and analysis are added, not only for \texttt{Libssh}, but also for a new HCS, namely \texttt{Libsoup}.

The prioritization strategy adopted is based on the number of features associated to a test case, assuming that the greater the number of changed features a test case covers in the system, the more fault-prone such a test case is. The idea is, after test selection, to rank the test cases in order to early execute those with a high probability of revealing faults. Considering the constraints of CI test budgets, early fault detection is essential because when a test case fails, test execution can be ended, and fewer resources are spent~\cite{PradoLimaTCPCIMapping}. 

The study we conducted to evaluate \texttt{FeaTestSelPrio}  has two main objectives. 
First, we evaluate the steps of \texttt{FeaTestSel} by conducting the same analysis as our previous work, but now by adding a new system. \texttt{FeaTestSel} results are compared with the results of a changed-file-oriented approach, regarding the percentage of test reduction and quality with respect to detected failures, using retest-all as baseline. We observe that the changed-file-oriented approach reaches a greater percentage but sacrifices failure-detection. \texttt{FeaTestSel} reaches a better trade-off considering these both factors and a cost-effective TCS.  In the analysis of the set of commits with logs, our feature-oriented selection reaches an average reduction in the number of test cases of $\approx$42\%. Also, on average, the execution time of \texttt{FeaTestSel} summed to execution time of the selected test cases fits well in a budget of 50\% of the average time required to execute all the tests available in the commit.

For the second objective of our evaluation, we study the applicability of \texttt{FeaTestSelPrio} by comparing the time it takes to select and prioritize test cases summed to the time to execute them. This total time is then compared with the time in between CI cycles. The average time to perform the selection and prioritization is 26.40 seconds. This time, when summed to the time to execute the selected test cases (worst case), is 119.75 seconds. The prioritization step of \texttt{FeaTestSelPrio} on average takes only 0.42 seconds. The prioritization step is also evaluated considering early-fault detection, leading to a reduction of 44\% in the budget required to detect a failure in comparison with a reduction of 23\% produced by the order generated by applying only \texttt{FeaTestSel}. In 86\% of the failed commits, the prioritization requires a lower budget.

In summary, this paper introduces a feature-oriented test case selection and prioritization approach to test HCSs, which has the main following contributions:
\begin{enumerate}
\item \texttt{FeaTestSelPrio} allows the selection and prioritization of test cases that are related to not only changed parts of the code, but to the entire feature changed in a commit. This provides a natural way to test HCS, considering their main building blocks;
\item  Differently from other approaches in the literature, our approach does not require historical data or training phase, requiring only the source code and test cases of the HCS to be applied;
\item \texttt{FeaTestSelPrio} contributes to reduce the number of test cases, without reducing efficacy in the number of revealed bugs. The prioritization results allow an early failure-detection and rapid feedback; and
\item  Our approach produces several intermediate outputs, including the traceability of test cases to source code, traceability of test cases to features, and system features and location of features in source code. These outputs can be used by software engineers for improving development and reducing testing costs.
\end{enumerate}

{The paper is structured as follows.  Section~\ref{RlWork} reviews related work. Section~\ref{motivating} contains an example that serves as motivation for a feature-oriented approach. Section~\ref{proposed} introduces the proposed approach and presents its implementation aspects. Section~\ref{methodology} describes the methodology adopted in the approach evaluation. Section~\ref{result} presents and analyses the obtained results.  Section~\ref{discussion} discusses some implications for research and practice, as well as} limitations our approach and threats to the validity of our results.  Section~\ref{conclu} concludes the paper.


\section{Related Work}
\label{RlWork}


In the literature, we can find three mapping studies that focus on reporting existing pieces of work on RT for HCSs~\cite{mendoncamap2022,Kumar2016,Runeson2012}. The proposed methods, approaches, or tools mainly focus on selecting a representative set of HCS variants to be tested, identify differences among products, and using the FM to derive and select test cases. More details of related work and observed limitations are presented in the following.

Some pieces of work are devoted to the selection/prioritization of the best product configurations to be tested, having as focus the FM ~\cite{Kumar2016,parejo2016multi,ensan11} and considering different goals such as:  combinatorial testing,  product similarity, coverage of variability and important features. These works do not directly deal with test cases, instead they focus on selecting a representative set of products. With a similar goal, some studies apply model-based test considering the delta concept. For instance, the work of Lity et al.~\cite{lity2019retest} captures commonality and variability of an evolving product line by means of differences between variants and versions of variants to select the test cases to be retested. Lachmann et al.~\cite{lachmann2015delta} introduce an incremental delta-oriented approach for improving Software Product Line (SPL) integration testing efficiency by prioritizing test cases for product variants. Al-Hajjaji et al.~\cite{al2017delta} selected the most dissimilar product to the previously tested ones, in terms of deltas. Lachmann et al.~\cite{lachmann2017risk} present a TCP approach based on risk-based testing, which can automatically compute component failure impact and component failure probabilities for each product variant under test.  

In the context of TCS, the approach proposed by Wang et al.~\cite{wang2017automated,wang2013automated} applies a classification of annotated test cases.  The idea is to ensure that all test cases associated with a specific functionality provided by the user are executed. 
Hajri et al.~\cite{hajri2020automating} present an automated test case classification and prioritization approach that supports use case-driven testing in product lines. These two approaches have a limited applicability, because they need the FM or other artifacts that represent variability as a source of information. This is a disadvantage, because they are not always available, and in the HCS evolution process  may be outdated~\cite{vierhauser2010flexible,ghanam2010linking,heider2012using}.
Other TCS approaches select a subset of existing test cases to be reused for testing a new product~\cite{lochau2012incremental,WANG2016,Xu2013}, also focusing more on the variants than in the test of the HCS as a whole.   

Jung et al.~\cite{jung2023automated} propose ActSPL, an automated method for reusing the existing test cases for a new product of a product family. The basic assumption is that, when a test case covers only the pieces of code commonly shared by two or more products, the test case is sharable for the products. By using this assumption,  ActSPL examines if a test case of an existing product covers pieces of code that belong to a new product. By doing so, ActSPL determines whether an existing test case is reusable for the new product. 
The work of Silveira Neto et al.~\cite{neto2010regression} describes an RT framework for HCSs at the integration level. The idea is to reduce testing effort by selecting and prioritizing test cases based on architectural similarities between products. However, it requires several input artifacts, which are often unavailable, such as test scripts and integration level test suites. In addition, the intervention of testing experts is necessary.  These approaches consider differences between two products, and not the whole product family. 
Another limitation is that they do not consider an evolution scenario.

Tufail et al.~\cite{Tufail17} present a systematic review on traceability techniques and tools that link test cases to requirements based on static information, without requiring the program execution or test coverage. But the pieces of works mentioned in the review do not deal with the concept of feature, the main HCS element.  Some pieces of work introduce methods based on changed files or versions~\cite{gligoric2015practical, bertolino2020learning, romano2018spiritus}. An example is the tool Ekstazi~\cite{gligoric2015practical} that calculates the checksum of the new file versions and considers that the file has changed if the checksums are different from the old file version ones. To select test cases, Ekstazi calculates the file dependencies of the test units. Bertolino et al.~\cite{bertolino2020learning} presents a TCS approach, applying a criterion based on static dependency analysis at the class level. These approaches can be used in the HCSs, but they work only for Java code and do not consider HCSs particularities. Some studies for HCSs that are also based on source code~\cite{meinicke2016essential, kim2012shared, kim2011reducing}, do not generate  traceability for features, but only link test cases to lines of code.  

The work of Tuglular and {\c{S}}ens{\"u}l{\"u}n~\cite{tuglular2019spl} generates a traceability between the feature and test cases, but using a specific language through annotations.  The code-based method of Jung et al.~\cite{jung2019automated} considers the similarity and variability of a product family, leaving out test cases unaffected by source code changes. But the evaluation considers only well-developed HCSs (i.e., toy systems), and test cases were developed specifically for the evaluation, what hampers the use of the approach in practice. 
In another work, Jung et al.~\cite{jung2020efficient} propose a TCS method to avoid repeating equivalent test runs that cover exactly the same source code sequence and produce the same test result on two or more variants. To identify equivalent test runs, test case execution traces and source code checksum values are used. The several steps involved may be quite costly if applied to large systems that are constantly evolving. Yet, it is specific for Java and both pieces of work do not consider traceability to features, but only to source code files.

Another limitation of existing studies is that some of them do not directly select or prioritize test cases and do not consider the SPL evolution, that is, the different versions of variants in an RT scenario.  For instance, the great majority does not consider particularities of the CI environment. 
TITAN~\cite{marijan2017titan} is a data-history  approach used to determine an optimal test order to ensure feature coverage, early fault detection, and reduced execution time. The approach considers that the test cases have macros indicating which HCS features they exercise. However, we can observe that in most open-source HCSs these macros do not exist. This hampers its applicability for general scenarios.  
Other studies of Marijan et al.~\cite{marijan2018practical, marijan2019learning} identify redundancy by analyzing the overlap of configurations options in a test set.  Afterward, the tests are classified as unique, fully redundant, or partially redundant.  Then, historical test information is used to determine which configurations have demonstrated high failure in previous test runs. Based on this information, the approach classifies partially redundant tests into effective and ineffective tests. The analysis uses code coverage per test case, and this dynamical strategy can degrade the performance of algorithms. Approaches that perform only static analysis are more suitable.

Prado Lima et al.~\cite{lima2020learning,prado2022cost} introduce two strategies to apply a TCP learning-based approach called COLEMAN in the CI of HCSs: the Variant Test Set Strategy (VTS) that relies on the test set specific for each variant; and the Whole Test Set Strategy (WST) that prioritizes the test set composed by the union of the test cases of all variants. COLEMAN is an approach that learns from the test case failure-history,  guided by a reward function. The main idea of these approaches is to deal with volatility of variants, that is, a new variant can be added in the cycle and new test cases can be added or removed. In the evaluation, WTS provides better results in the less restrictive budgets, and VTS the opposite. WTS seems to better mitigate the problem of beginning without knowledge, and is more suitable when a new variant to be tested is added. Our work differs from those because we first select a test set based on the evolution of features, which contributes to reduce costs and bypass the variant volatility problem. 

In summary, existing work present the following main limitations: (i)  are dependent on models or other artifacts  (e.g., FM or HCS architecture) that may not exist or be outdated~\cite{lity2019retest, lachmann2015delta, al2017delta, lachmann2017risk, neto2010regression, wang2017automated, wang2013automated, hajri2020automating}; (ii) require a failure-history or dynamic analysis~\cite{marijan2017titan,marijan2018practical,marijan2019learning,prado2022cost,lima2020learning}, what is costly and may be not suitable for a CI scenario; (iii) do not consider HCS particularities and/or languages such as C and C++, largely adopted for the HCS development~\cite{gligoric2015practical, bertolino2020learning, romano2018spiritus,tuglular2019spl}; (iv)~consider that there is a kind of mapping from code to the test cases or that the HCSs are developed following a specific format~\cite{marijan2017titan,jung2019automated}; and (v) do not work with the concept of features, which are fundamental units of design and communication in the HCS context~\cite{meinicke2016essential, kim2012shared, kim2011reducing,jung2020efficient}.
Our approach, presented in Section~\ref{proposed}, addresses these limitations. We aim for an approach that has automated TCS and TCP based on features. It relies only on the static analysis of the source code, and works for HCSs written in C/C++ language. Moreover, we introduce a prioritization step to deal properly with the test budgets.


\section{Motivating Example}

\label{motivating}

In this section, we present a motivating example showing the importance of considering the relation between features and test cases for RT during evolution of HCSs. 
To this end,  we use the system \texttt{Libssh} (presented in details in Section~\ref{system}), which is in constant evolution, updated by several developers, sometimes more than once a day. 
This is the common case of open-source HCSs, leading to the issues discussed below.

By analyzing the CI cycles and the test case evolution in \texttt{Libssh}, we observe that, in the great majority of the commits, the retest-all technique is executed after each cycle.
In this scenario, suppose that a developer needs to make a small change in the system related to a feature of this HCS, and after this change the retest-all technique is adopted. This is an expensive and time-consuming approach that uses additional computation resources for every change, even the simple ones. This retest may be necessary several times due to the number of developers involved.\footnote{\texttt{Libssh} is developed by the collaboration among 110 developers, according to \url{https://github.com/libssh/libssh-mirror/graphs/contributors}} 

\begin{figure}[!tp]
  \centering
  \includegraphics[width=.70\linewidth]{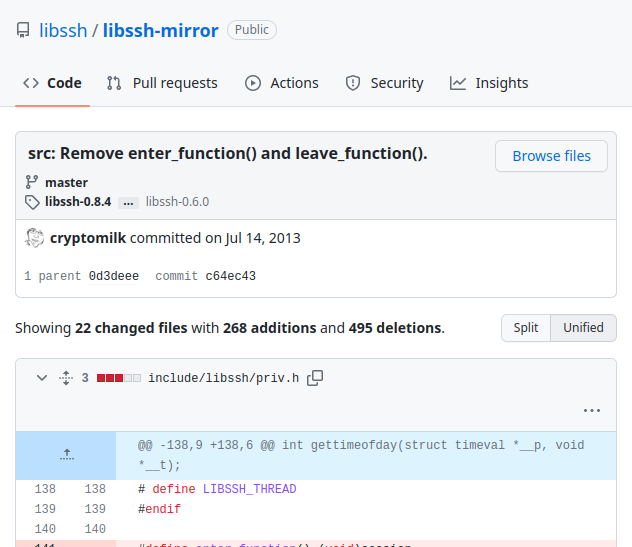}
  \caption{Number of modifications made in commit \texttt{c64ec43}}
  \label{fig:commit_c64ec43}
\end{figure}

For illustration, we take as example the  commit \texttt{c64ec43},\footnote{\url{ https://gitlab.com/libssh/libssh-mirror/-/commit/c64ec43}} which performs the removal of the functions \texttt{enter\_function()} and \texttt{leave\_function()}. This modification is related to 22 changed files with 268 additions and 495 deletions, as shown in Figure~\ref{fig:commit_c64ec43}. In this commit, 117 test cases are available, making the retest-all technique to take up to five minutes to run per variant.\footnote{Time per variant is found at \url{https://gitlab.com/libssh/libssh-mirror/-/pipelines}} Notice here that the HCS can receive several updates a day, as can be seen in the repository, and many variants must be tested.

\begin{figure}[!tp]
  \centering
  \includegraphics[width=.9\linewidth]{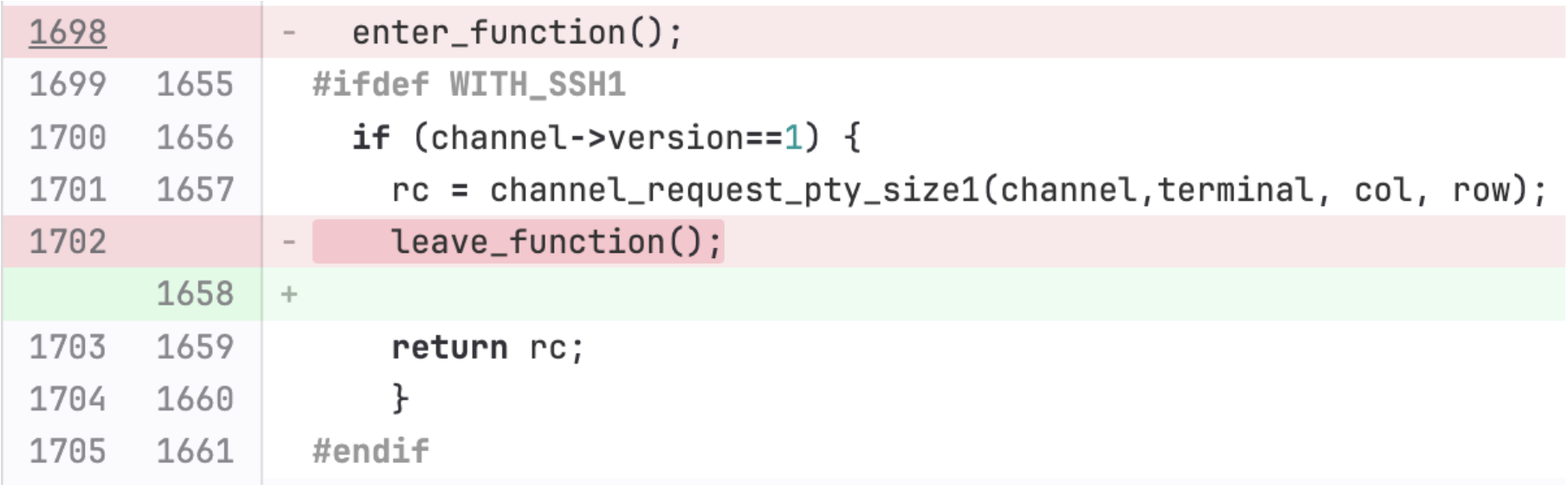}
  \caption{Example of modification in commit \texttt{c64ec43}}
  \label{fig:motexample}
\end{figure}

Instead of always executing all test cases (i.e., retest-all), we can adopt a smarter strategy. An alternative is to trace test cases to the 22 files changed and select only the test cases associated to them. But this technique may not select the test cases related to all the features changed in the commit, and that can cause failures in unchanged files because of features interactions \cite{meinicke2016essential} or the Ripple effect~\cite{yau1978ripple}.  An example is presented in Figure~\ref{fig:motexample}, in which the feature \texttt{WITH\_SSH1} was changed in the commit referred above. 
In this specific change, the call to \texttt{leave\_function()} was excluded. However, the exclusion of this call can change all the functionality of \texttt{WITH\_SSH1} and impact other unchanged files that also implement this feature, such as \texttt{options.c} and \texttt{channels1.c}.

While analyzing the execution of test cases for \texttt{Libssh}, we also observed that the test cases are always executed in the same order. Currently, CI cycles do not have any strategy or method to prioritize the test cases based on the information about \texttt{Libssh} evolution. Thus, test cases are executed without considering they have different importance. For instance, some test cases have higher probability of failing than others. Then, to produce a rapid feedback for the test, it is important these test cases are executed first. A solution to  this is to rank the test cases. To address this issue, we introduce a feature-oriented selection and prioritization approach (presented in the next section). This approach is capable of capturing the relationship between features and test cases, including all impacted test cases in the selected test set.

From the \texttt{Libssh} repository, we obtained insights that motivated our work.
For instance, we can observe that in the commit \texttt{d7477dc7}\footnote{\url{ https://gitlab.com/libssh/libssh-mirror/-/commit/d7477dc7}} there are 179 test cases, from which 4 failed during the test. In a more detailed analysis, we see that these test cases cover on average 6.5 features, while the test cases that not failed cover 3.9. This serves as motivation to our approach, which generates a rank based on feature coverage, considering  test cases that are related to a greater number of features are more fault-prone.


\section{Proposed Approach}
\label{proposed}

\begin{figure}[!tp]
  \centering
 \includegraphics[width=0.6\linewidth]{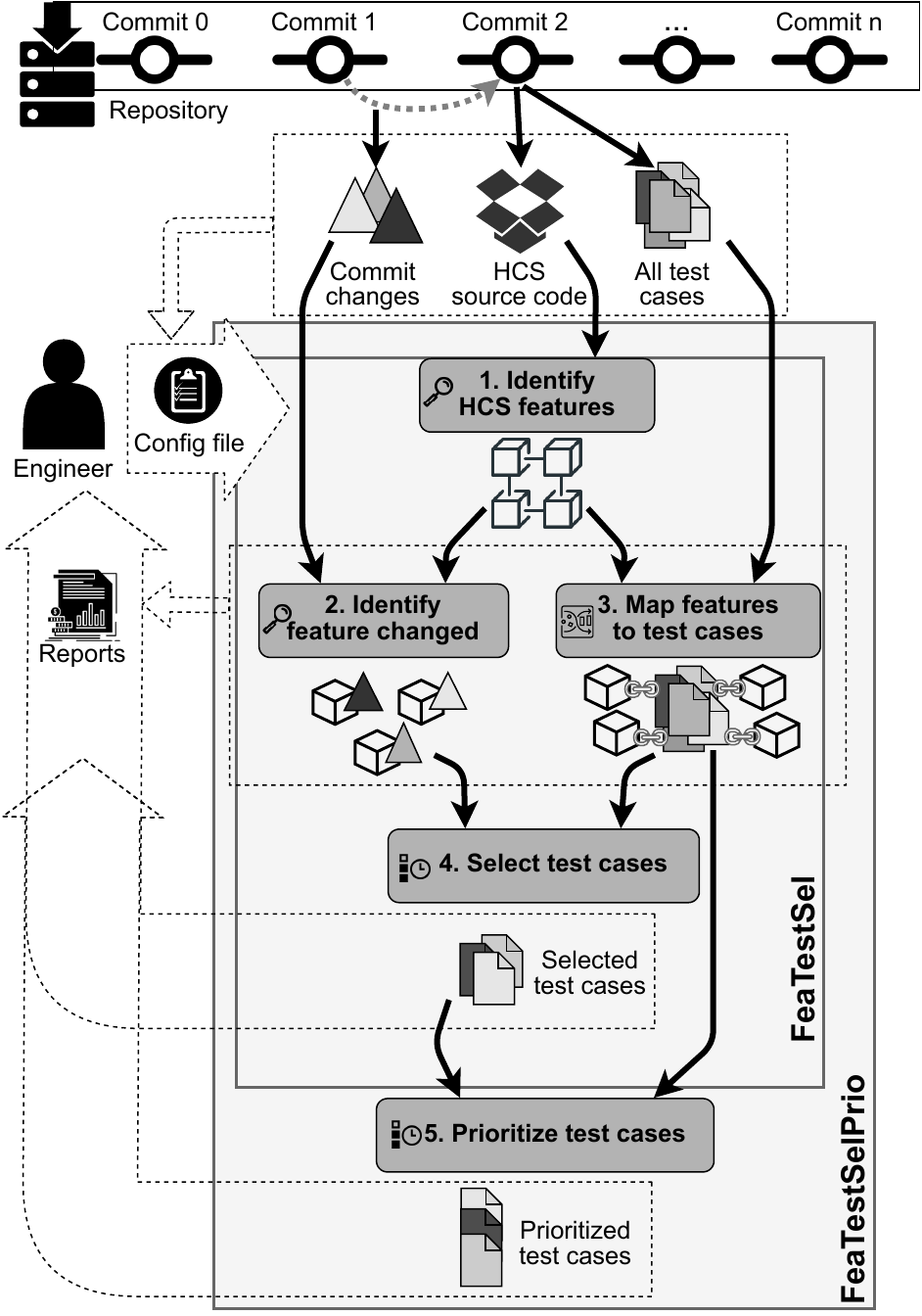}
  \caption{Overview of our approach}
  \label{fig:ApproachStep}
\end{figure}

In our previous work, we introduced  {\tt FeaTestSel} ({\bf Fea}ture-oriented {\bf Test} Case {\bf Sel}ection for Highly Configuration Systems)~\cite{mendoncaTCS2023}, which consists of four steps that are presented in Figure~\ref{fig:ApproachStep}.  The tester needs to only provide a configuration file ({\it Config file}) containing the paths to the source code of the HCS and the test case folder. In Step 1, {\it Identify HCS features}, the source code corresponding to each feature of the HCS is determined. The lines of code that implement each feature of the system are identified automatically based on pre-processor directives. After this, two independent steps are performed. In Step 2, {\it Identify features changed}, the source code of the current commit is compared with the previous one to identify feature changes (i.e., features modified, added, or removed). In Step 3, {\it Map features to test cases}, the lines of code exercised  by each test case are identified and  trace links between feature and test cases are created. In the fourth step, {\it Select test cases}, the output of Steps~2 and~3 are used  to select test cases related to feature changes in a given commit. The main \textit{output} consists of the selected test cases and reports with traceability information between test cases and features.

\sloppy
In the present work, we devise an extension to \texttt{FeaTestSel}, called \texttt{FeaTestSelPrio} ({\bf Fea}ture-oriented {\bf Test} Case {\bf Sel}ection and {\bf Pri}oritization for Highly Configurable Systems). This extension includes an additional step, {\it  Prioritize test cases}, that performs the test case prioritization  considering the reports generated by  \texttt{FeaTestSel}. 
We can observe in Figure~\ref{fig:ApproachStep} that this step is applied considering as input the test cases selected in Step 4 of \texttt{FeaTestSel}, combining selection and prioritization techniques. 
In addition to the test cases, the prioritization step also uses as input the traceability of features to test cases generated  in Step 3. At the end, a set of prioritized test cases is produced.

\texttt{FeaTestSelPrio} is designed to be lightweight; it does not need any learning process or any long history of changes or test failures to perform the test case selection and prioritization. Thus, our approach can be executed after each commit,  identifying feature implementations, feature changes, and feature to test traceability using the latest version of the HCS. 
Our approach was implemented in Python, 
version 3.9.10. The outputs are saved as CSV files, for which we adopt PANDAS.\footnote{\url{https://pandas.pydata.org/}}
In the next subsections, we describe each step of our approach in details and present more implementations aspects.

\subsection{Input}

\begin{figure}[!htp]
  \centering
 \includegraphics[width=.8\linewidth]{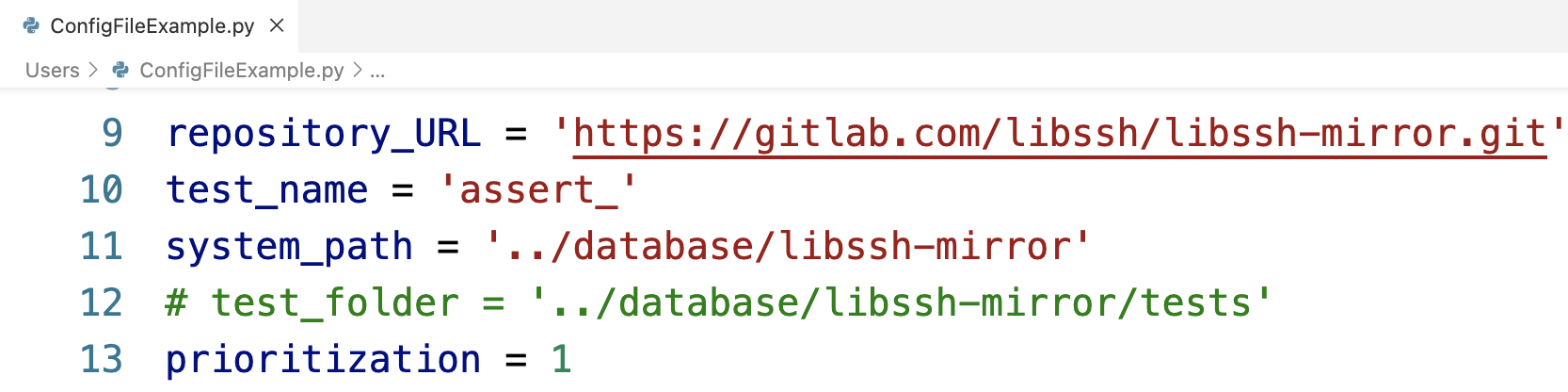}
  \caption{Configuration file used as input}
  \label{fig:scriptimg}
\end{figure}

As input, \texttt{FeaTestSelPrio} receives a configuration file 
 ({\it Config file}) containing  paths to some folders used as source of information, as illustrated in Figure~\ref{fig:scriptimg} for the system \texttt{Libssh}. 
They are: (i)~\texttt{repository\_URL}: contains the URL to the repository of the HCS (e.g., the URL of the HCS on GitHub); (ii) \texttt{system\_path}: indicates the path of the source code folder with the implementation of the features;  (iii) \texttt{test\_name}: defines a pattern in the nomenclature of test cases that allows the approach to identify which source code files are related to test cases. Alternatively, the software engineer can provide the folder name,  \texttt{test\_folder}, used to store test cases, when this is a practice in the project. In this case, the test case selection will perform faster.  However, using the test names brings the benefit that all system files will be checked, since by using a good search string, hardly any test case file will be forgotten; and (iv)~\texttt{prioritization}: contains 1 whether the prioritization will be performed after the selection, or 0 otherwise.

\subsection{Identify HCS Features}

To mine features and their changes, our implementation abstracts the source code of a commit snapshot at the level of pre-processor directives, which are  distinguished in conditional blocks (i.e., \texttt{\#if}, \texttt{\#ifdef}, \texttt{\#elif}, \texttt{\#else}, and \texttt{\#ifndef}), definition lines (i.e., \texttt{\#define} and \texttt{\#undef} directives), or import file lines containing \texttt{\#include} directives, similarly to related work~\cite{MichelonGPCE2021}. This abstraction is less computationally expensive, as we do not need to analyze the abstract syntax tree to obtain, for each file, the lines of code containing pre-processor directives.  The approach uses a Constraint Satisfaction Problem Solver~\cite{Schiex2019} to reliably identify features interacting/depending on the execution of other features~\cite{Benavides2006}, and thus which features belong to which conditional blocks to obtain the features lines.

The code excerpt in Figure~\ref{lst:runningexample} is used to illustrate the strategy adopted for mining features lines. The figure contains four variation points (i.e., conditional blocks) wrapped by conditional directives. For each conditional block, the approach creates constraints related to each particular conditional block of code. For instance, lines~1-3 belong to feature \texttt{A}. This is the simplest case, where there are no interactions or nested features. But the block of code of lines~6-8 belongs to a feature internally defined, inside feature \texttt{A}. This block of code of lines~6-8 is activated when feature \texttt{A} is selected, and thus when \texttt{B} is greater than 10. However, this is not the only constraint to take into account to determine which features belong to the block of code of lines~6-8 because there is an outermost block wrapping  lines~6-8 with the conditional expression \texttt{\#if C}. In this case, feature \texttt{C} also has to be selected so that this block of code can be executed. Therefore, in such cases, where multiple features imply executing a block of code, a heuristic is adopted to consider the lines as part of the closest feature (not defined internally via \texttt{\#define} directive) to the block of code. In this way, the block of code of lines~6-8 is assigned to the feature \texttt{C}. Therefore, the lines of code of feature \texttt{C} begins at line~5 and ends at line~9.
 
We also have a corner case example~\cite{Ludwig2019} at lines~11-13, where there is a negated conditional expression, namely the block of code is executed when feature \texttt{D} is not selected. In this case, there are no nested features or feature interactions, and this block of code is thus considered part of the system core (\texttt{BASE} feature), as there is no feature responsible for executing this block.
After getting the features responsible to execute each block of code, we obtain the line numbers of each file that belongs to a feature. Therefore, lines~\mbox{1-3} are related to feature \texttt{A}, lines~5-9 to feature \texttt{C}, lines~11-13 to feature \texttt{D}, and to \texttt{BASE}, as well as line~15, which is outside any variation point.

 \begin{figure}[!t]
 \centering
 {\footnotesize
\begin{lstlisting}[escapechar=\%,language=C,basicstyle=\footnotesize,numbersep=-3pt,frame=single, xleftmargin=.20\columnwidth, xrightmargin=.20\columnwidth, ,frame=shadowbox, rulesepcolor=\color{gray}]
 1    #ifdef A
 2        #define B 15
 3    #endif
 4
 5    #if C
 6        #if B > 10
 7           <code>
 8        #endif
 9    #endif
10
11    #ifndef D
12           <code>
13    #endif
14
15    <code>
\end{lstlisting}}
\caption{Conditional blocks of feature implementations.} 
\label{lst:runningexample}
 \end{figure}

\subsection{Identify Feature Changed}

The outputs of previous steps are used to identify the features and their locations. When observing differences between commits, a feature change is identified in two cases: (i) a new feature is found, i.e., a new pre-processor directive with a feature annotation is added; or (ii) a change in an existing feature occurs, i.e., a change in between pre-processor directives that delimit blocks of code belonging to a feature. To obtain this information, we first collect all conditional block macros and all \texttt{\#defines} present in all files from each release commit. Then, we look for macros  never defined within the source code, i.e., that can only be defined externally by the user, from the command line. In this way, we obtain the macros that can be considered as features of the system. 

For each Git commit $n$, we generate the differences between the actual commit and the previous commit $n-1$. In the case of the initial Git commit in the project, we deal with all inserted files as the point of difference. From these differences, we can obtain the tree node that reflects the changes. If there are modifications to any external features or discrepancies in non-code files, such as binary files, BASE is considered as the altered feature. In other words, for any code additions or removals within the project body that do not pertain to an external feature, the root feature BASE is identified as the modified node.

In summary, after knowing all blocks of code that belong to the features, our approach uses Git diff\footnote{\url{https://git-scm.com/docs/git-diff}} to collect code fragments that differ for the same file from one commit to another. This process obtains the differences of fragments with patches. These patches represent the differences between two text files in a line-oriented manner, as calculated by a diff utils library.\footnote{\url{https://java-diff-utils.github.io/java-diff-utils/}}

\subsection{Map Features to Test Cases}

To identify dependencies between test cases and blocks of code belonging to features, this step uses static analysis and the tool Test2Feature~\cite{Test2FeatureSPLC2022}. First, a dependency graph using all code available in the repository is created. Then,  the dependencies between test cases and source code implementing features are collected. Finally, using the output of Step 1, namely the lines of code implementing each feature, the approach creates trace links between the test cases and the features they are related to. 

Basically, a merge between the output of Step 1 and the test cases found in the HCS (i.e., links between the location of the features along with the location of the test cases) is performed. In this way, it is possible to know exactly the location of the tests and features per line of the files. Initially, a merge is performed considering the localization files, then, a filter is applied  considering the location of the code lines. The output of this step is stored in a CSV file.

This step of our approach is implemented based on Doxygen.\footnote{\url{https://doxygen.nl/index.html}} Doxygen supports visualization of the relationships between various elements  through  dependency graphs, inheritance, and collaboration diagrams, generated  automatically, in different formats. Our implementation uses Doxygen for C/C++ and  the function dependency graph in XML format. This tool generates the dependency graph as XML files, performing static analysis of the source code. 
Doxygen for C/C++ properly parses pre-processor directives by default, considering this is a common construct of C/C++. We defined the minimal set of parameters in order to generate all possible dependencies available in Doxygen and to make the tool executes faster.

\subsection{Select Test Cases}

This step selects test cases that cover the features  changed in the commit (output of Step~2: \textit{Identify feature changed}). When a change in a feature is identified, all test cases that are linked to this specific feature are selected. We consider a test case covers a feature when it exercises the lines within the blocks of code belonging to a feature, leveraging the trace links created in Step~3 (\textit{Map features to test cases}).

Despite its simplicity, this selection has an advantage when compared to existing approaches. Approaches that are changed-oriented mostly select test cases that are direct related to the changed files, without considering features. In our case, even if the changed file is not touched by the test case, but the test case touches other parts of the features being changed in the given commit, that test case is selected. Since features are building blocks of HCSs, it is important to verify the behavior of the changed features, and it is also important to consider the coverage of the features.

\subsection{Prioritize Test Cases}
To perform this step, the parameter \texttt{prioritization} in the configuration file must be set to 1. In such a case, the test case prioritization based on features is performed using the trace links between features and test cases (Step~3). The main idea is that test cases associated to a greater number of features are ranked first, because they have a higher probability to reveal faults, as shown in our motivating example. Also, these test cases that are linked to different features may exercise feature interactions, which a common HCS characteristic \cite{meinicke2016essential}.

To implement this prioritization strategy to maximize the number of features tested, we create a list for each test case, indicating the features it is associated. We then generate an overall coverage list that includes the names of the test cases, the list of features they cover, and a count of the number of features covered by each test case.  The greater the number of features a test case covers in the system, the higher its priority. Test cases covering more features are  ranked first. In case of a tie in the number of covered features, the tiebreaker is random.



\subsection{Illustrative Example}
\label{validation}

\sloppy
To illustrate the required input and produced outputs of \texttt{FeaTestSelPrio}, we use again \texttt{Libssh} and focus on the commit~\texttt{c64ec43}.\footnote{\url{https://github.com/libssh/libssh-mirror/commit/c64ec43}} This commit was chosen because it represents well the evolution of an HCS, with changes occurring in five different features: \texttt{WITH\_ZLIB}, \texttt{WITH\_SSH1}, \texttt{WITH\_SFTP}, \texttt{WITH\_SERVER}, and \texttt{BASE}. The feature \texttt{BASE} encompasses the implementation of all parts of the HCS that do not belong to a feature (i.e., blocks of code that are not wrapped in pre-processor directives). \texttt{BASE} corresponds to a large portion of code, and mostly changes in all commits.

\begin{figure}[!htp]
  \centering
  \includegraphics[width=.6\linewidth]{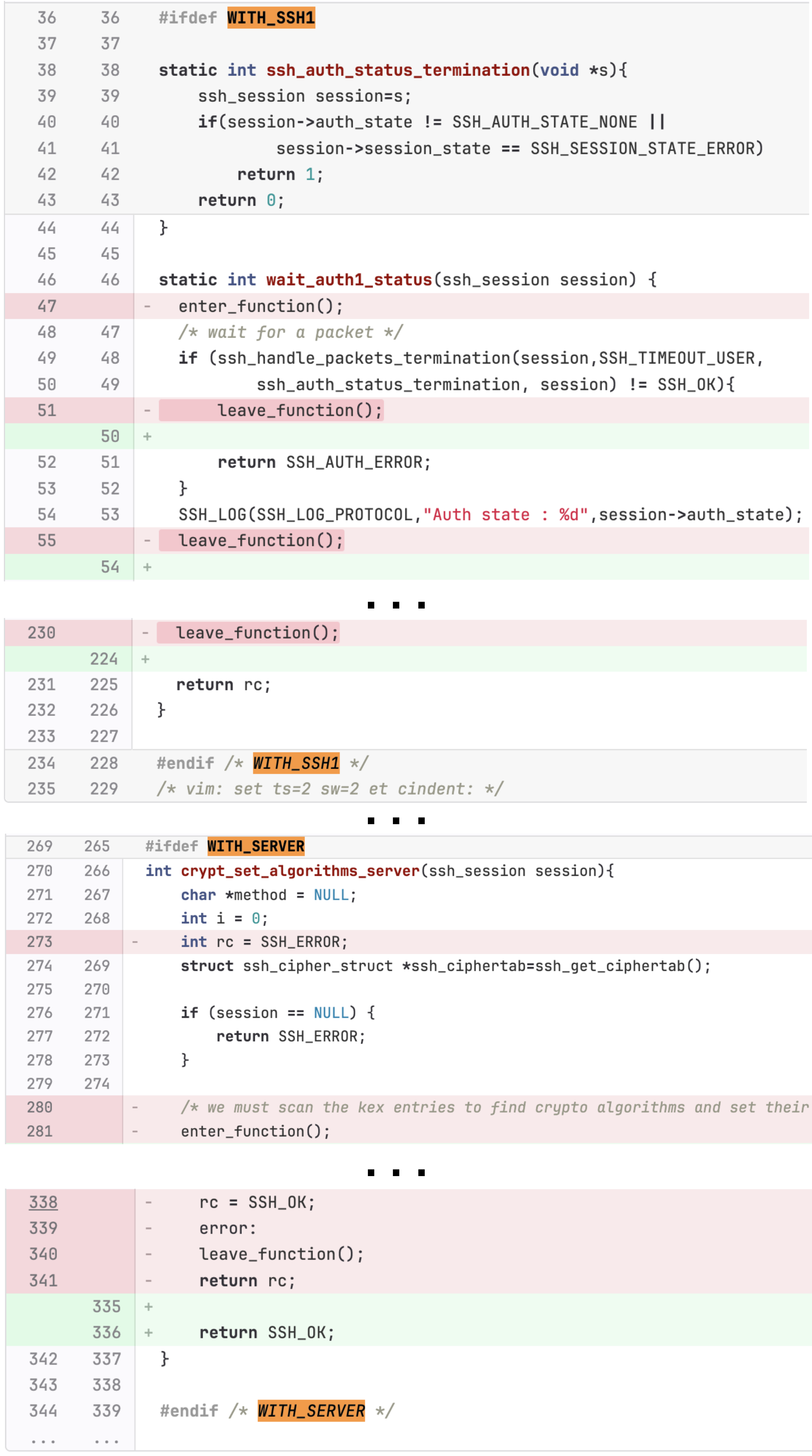}
  \caption{Excerpt of source code for \texttt{Libssh\$} features changed as part of commit \texttt{c64ec43}}
  \label{fig:featChanged2}
\end{figure}

\begin{figure}[!htp]
  \centering
  \includegraphics[width=.8\linewidth]{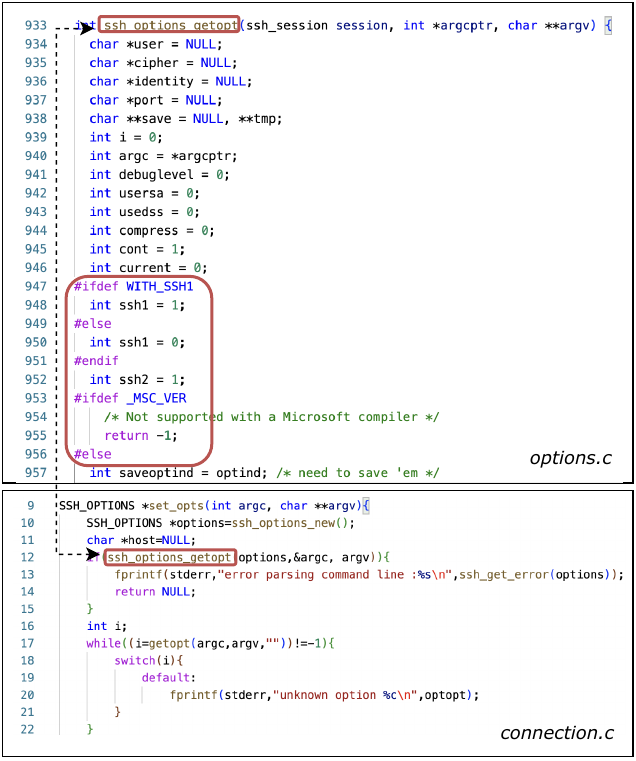}
  \caption{Traceability between test cases and feature}
  \label{fig:testline}
\end{figure}

Figure~\ref{fig:featChanged2} presents an excerpt of the source code corresponding to some  \texttt{Libssh} features. We can observe changes in the feature \texttt{WITH\_SSH1}, on lines 47, 51, and 55; and in the feature \texttt{WITH\_SERVER}, on lines 273 and 281.  
In our example, consider the test case \texttt{set\_ops} from the file \texttt{connection.c}, presented in Figure~\ref{fig:testline}. From this test case, Table~\ref{tab:testtoline} presents an excerpt of the CSV document generated by our approach.  The test case \texttt{set\_ops} has traceability to the files \texttt{sample.c}, \texttt{error.c} and \texttt{options.c}. Moreover,  it is possible to trace specific functions and lines. Figure \ref{fig:testline} presents an example of the traceability between the test file \texttt{connection.c}  and the file \texttt{options.c}, showing the traceability at the function and feature levels.
In Table~\ref{tab:testtoline} we can see that the function \texttt{ssh\_options\_getopt}  corresponds to the lines 933 to 1102  in the test file \texttt{connection.c}, where there is a call to this function on line 12. Thus,  there is a  traceability between the test case \texttt{set\_opts} and the function \texttt{ssh\_options\_getopt}.

To illustrate feature traceability, we also use the granularity of lines of code and the feature \texttt{WITH\_SSH1} as an example. Table \ref{tab:featuretoline} presents an excerpt of the \texttt{CSV} file generated by our approach. We can see that \texttt{WITH\_SSH1} is found between lines 947 to 949 in the file \texttt{options.c}. Thus, to know exactly which test cases touch which feature, we use the two CSV files, performing a merge of the related lines of code. We can see in  Figure \ref{fig:testline}  the \texttt{\#ifdef} corresponding to the feature \texttt{WITH\_SSH1} and check this feature is inside the function \texttt{ssh\_options\_getopt} and the test case \texttt{set\_opts} touches this function. Then, in the case of a change in this feature, this test case must be selected. In addition to this,  we can observe in Figure~\ref{fig:testline} the feature is between the range of the function \texttt{ssh\_options\_getopt}  that covers lines 933 to 1102. Table~\ref{tab:testSelection} presents the CSV file generated at the end of Step 4 where we can see in the test case \texttt{set\_opts}.

\begin{table}[!tp]
\caption{Traceability of Test Cases to Source Code Lines}
\label{tab:testtoline}
\resizebox{1\linewidth}{!}{
\begin{tabular}{l|l|l|l|c|c}
\hline
\textbf{Test File} & \textbf{Test Case} & \textbf{Target File} & \textbf{Target Function} & \textbf{Line From} & \textbf{Line To} \\ 
\hline
\hline
connection.c                           & set\_opts                              & sample.c                                 & host                                         & 39                                     &  39                                   \\ \hline
connection.c                           & set\_opts                              & error.c                                  & ssh\_get\_error                              & 109                                    & 113                                 \\ \hline
connection.c                           & set\_opts                              & options.c                                & ssh\_options\_getopt                         & 933                                    & 1102                                \\ \hline
\end{tabular}}
\end{table}

\begin{table}[!ht]
	\caption{Traceability of Test Cases to Features}
 \centering
	\label{tab:featuretoline}
\small
\begin{tabular}{l|l|c|c}
\hline
\multicolumn{1}{c|}{\textbf{Target File}} & \multicolumn{1}{c|}{\textbf{Feature Name}} & \multicolumn{1}{c|}{\textbf{Feat From}} & \multicolumn{1}{c}{\textbf{Feat To}} \\ 
\hline
\hline
src/client.c                             & WITH\_SSH1                                & 340                                    & 343                                 \\ \hline
src/channels.c                           & WITH\_SSH1                                & 1292                                   & 1298                                \\ \hline
src/channels.c                           & WITH\_SSH1                                & 1655                                   & 1661                                \\ \hline
src/options.c                            & WITH\_SSH1                                & 947                                    & 949                                 \\ \hline
src/channels1.c                          & WITH\_SSH1                                & 41                                     & 389                                 \\ \hline
src/server.c                             & WITH\_SSH1                                & 344                                    & 348                                 \\ \hline
\end{tabular}
\end{table}

\begin{table*}[!htp]
	\caption{Partial Set of Test Cases Selected by FeaTestSel}
	\label{tab:testSelection}
	\centering
	\resizebox{1\textwidth}{!}{
     \fontsize{14}{14}\selectfont
\begin{tabular}{l|l|l|l|c|c|l|c|c}
\hline
\multicolumn{1}{c|}{\textbf{Test File}} & \multicolumn{1}{c|}{\textbf{Test Case}} & \multicolumn{1}{c|}{\textbf{Target}} & \multicolumn{1}{c|}{\textbf{Target Function}} & \multicolumn{1}{c|}{\textbf{Line}} & \multicolumn{1}{c|}{\textbf{Line}} & \multicolumn{1}{c|}{\textbf{Feature}} & \multicolumn{1}{c|}{\textbf{Feat}} & \multicolumn{1}{c}{\textbf{Feat}} \\ 

 &  & \multicolumn{1}{c|}{\textbf{File}} & & \multicolumn{1}{c|}{\textbf{From}} & \multicolumn{1}{c|}{\textbf{To}} & \multicolumn{1}{c|}{\textbf{Name}} & \multicolumn{1}{c|}{\textbf{From}} & \multicolumn{1}{c}{\textbf{To}} \\ 
\hline
\hline
connection.c                           & set\_opts                              & options.c                                & ssh\_options\_getopt                         & 933                                    & 1102                                 & WITH\_SSH1                                & 947                                    & 949                                 \\ \hline
test\_exec.c                           & do\_connect                            & channels.c                               & ssh\_channel\_open\_session                  & 920                                    & 936                                  & WITH\_SSH1                                & 925                                    & 929                                 \\ \hline
test\_exec.c                           & do\_connect                            & channels.c                               & ssh\_channel\_request\_exec                  & 2382                                   & 2428                                 & WITH\_SSH1                                & 2395                                   & 2399                                \\ \hline
bench\_raw.c                           & upload\_script                         & channels.c                               & ssh\_channel\_open\_session                  & 920                                    & 936                                  & WITH\_SSH1                                & 925                                    & 929                                 \\ \hline
bench\_raw.c                           & upload\_script                         & channels.c                               & ssh\_channel\_request\_exec                  & 2382                                   & 2428                                 & WITH\_SSH1                                & 2395                                   & 2399                                \\ \hline
bench\_raw.c                           & benchmarks\_raw\_up                    & channels.c                               & ssh\_channel\_open\_session                  & 920                                    & 936                                  & WITH\_SSH1                                & 925                                    & 929                                 \\ \hline
bench\_raw.c                           & benchmarks\_raw\_up                    & channels.c                               & ssh\_channel\_request\_exec                  & 2382                                   & 2428                                 & WITH\_SSH1                                & 2395                                   & 2399                                \\ \hline
bench\_raw.c                           & benchmarks\_raw\_down                  & channels.c                               & ssh\_channel\_open\_session                  & 920                                    & 936                                  & WITH\_SSH1                                & 925                                    & 929                                 \\ \hline
\end{tabular}
}
\end{table*}

In Step 5, {\it Prioritize Test Case},  we count the number of features covered by the test cases. Using Step 3 of {\tt FeaTestSel}, which maps features to test cases, we can calculate this coverage, updating it with each evolution of the HCS, i.e., commit by commit. Table \ref{tab:featCov} provides a partial record of the results of this strategy. In this case, it can be observed that, in the first three rows, the test cases cover exactly eight features. In the range from line four to ten, test cases cover three features. The next test cases in the rank are not presented in the table, but cover a lower number of features. The test cases ranked in the end of the list do not cover any feature.

\begin{table*}[!htp]
	\caption{Example - Feature Coverage}
	\label{tab:featCov}
	\centering
	\resizebox{1\textwidth}{!}{
 \fontsize{14}{14}\selectfont
\begin{tabular}{c|l|l|p{6.5cm}|c}
\hline
\multicolumn{1}{c|}{\textbf{Index}} & \multicolumn{1}{c|}{\textbf{TestFile}} & \multicolumn{1}{c|}{\textbf{TestCase}} & \multicolumn{1}{c}{\textbf{FeatureNameList}} & \multicolumn{1}{|c}{\textbf{Count}} \\
\hline
\hline
1 & torture\_pki.c        & torture\_pki\_generate\_key\_dsa            & {[}'WITH\_SSH1', 'WITH\_PCAP', 'HAVE\_ECC', 'HAVE\_OPENSSL\_ECC', 'HAVE\_LIBCRYPTO', 'DEBUG\_CRYPTO', 'BASE', 'HAVE\_LIBGCRYPT'{]} & 8 \\ \hline 
2 & torture\_pki.c        & torture\_pki\_generate\_key\_rsa            & {[}'HAVE\_ECC', 'WITH\_SSH1', 'HAVE\_LIBGCRYPT', 'WITH\_PCAP', 'HAVE\_LIBCRYPTO', 'BASE', 'DEBUG\_CRYPTO', 'HAVE\_OPENSSL\_ECC'{]} & 8 \\ \hline
3 & torture\_pki.c        & torture\_pki\_generate\_key\_rsa1           & {[}'WITH\_PCAP', 'BASE', 'HAVE\_LIBCRYPTO', 'DEBUG\_CRYPTO', 'WITH\_SSH1', 'HAVE\_ECC', 'HAVE\_LIBGCRYPT', 'HAVE\_OPENSSL\_ECC'{]} & 8 \\ \hline
4 & connection.c          & set\_opts                                   & {[}'WITH\_SSH1', 'BASE', '\_MSC\_VER'{]}                                                                                           & 3 \\ \hline
5 & test\_socket.c        & main                                        & {[}'WITH\_SSH1', '\_WIN32', 'BASE'{]}                                                                                              & 3 \\ \hline
6 & torture.c             & torture\_ssh\_session                       & {[}'WITH\_SSH1', 'WITH\_PCAP', 'BASE'{]}                                                                                           & 3 \\ \hline
7 & torture\_keyfiles.c   & torture\_privatekey\_from\_file             & {[}'HAVE\_LIBGCRYPT', 'HAVE\_LIBCRYPTO', 'BASE'{]}                                                                                 & 3 \\ \hline
8 & torture\_keyfiles.c   & torture\_privatekey\_from\_file\_passphrase & {[}'BASE', 'HAVE\_LIBCRYPTO', 'HAVE\_LIBGCRYPT'{]}                                                                                 & 3 \\ \hline
9 & torture\_keyfiles.c   & torture\_pubkey\_generate\_from\_privkey    & {[}'HAVE\_LIBCRYPTO', 'HAVE\_LIBGCRYPT', 'BASE'{]}                                                                                 & 3 \\ \hline
10 & torture\_knownhosts.c & torture\_knownhosts\_port                   & {[}'WITH\_PCAP', 'BASE', 'WITH\_SSH1'{]}                                                                                           & 3 \\ \hline
\multicolumn{1}{c|}{\textbf{..}} & \multicolumn{1}{c|}{\textbf{..}}  & \multicolumn{1}{c|}{\textbf{...}} & \multicolumn{1}{c|}{\textbf{...}} & \multicolumn{1}{c}{\textbf{...}}    \\ \hline
\end{tabular}}
\end{table*}

\section{Evaluation Setup}
\label{methodology}

This section presents details of the evaluation of \texttt{FeaTestSelPrio}. The main goal is to evaluate the applicability of our approach and compare their results with a changed-file-oriented approach, having the retest-all technique as baseline.

\subsection{Research Questions}
\label{RQs}
Based on the evaluation goal, we formulated four {\it Research Questions (RQ)}:

\vspace{0.1cm}\noindent \textbf{RQ1:} \textit{How is the performance of \texttt{FeaTestSel} when compared to the performance of the retest-all and changed-file-oriented approaches?} This RQ focuses on the test case selection. We compare \texttt{FeaTestSel} with a changed-file-oriented approach, which selects test cases associated to the files changed in the commit (see Section~\ref{RlWork}). The analysis also considers the time taken to perform the selection and the percentage of reduction in the number of test cases in comparison to executing all the test cases available for the commit (retest-all technique).

\vspace{0.1cm}\noindent \textbf{RQ2:} \textit{What is the failure-detection quality of the test cases when \texttt{FeaTestSel} is used?} This question investigates whether by reducing the number of test cases, it is still possible to maintain the test quality in terms of detected failures. We then analyze  the number of failures detected by the test cases selected by \texttt{FeaTestSel} and by the changed-file-oriented approach in comparison with the number of failures detected when all test cases available are executed (retest-all technique).

\vspace{0.1cm} \noindent \textbf{RQ3:} \textit{Is \texttt{FeaTestSel} applicable  considering budget constraints of industrial HCSs?} 
This question aims to assess the applicability of our test case selection approach by analyzing the time between CI cycles, with the execution time of the selected test cases summed to the time spent to perform the selection. We want to check whether the execution of the selected test set generated by our approach implies in a reduced time to execute the tests. Moreover, we are interested in the time our approach takes to execute. If such time is too long, the use of \texttt {FeaTestSel} is impracticable in a CI environment. 

\vspace{0.1cm} \noindent \textbf{RQ4:} \textit{What is the performance of \texttt{FeaTestSelPrio} for early fault detection?} In this RQ, we evaluate the performance of the prioritization step of our approach. Using an indicator for early fault detection, we compared the test case order  produced by \texttt{FeaTestSelPrio} with the order produced by \texttt{FeaTestSel}. In fact, \texttt{FeaTestSel} does not produce any order, but the result  of our implementation is given according to the order of the test cases appear in the directory. The order in which files are returned depends on the specific implementation of the underlying file system. However, the order is generally alphanumeric, that is, the files are returned in lexicographic order. Moreover, we also evaluated the execution time for the prioritization step of \texttt{FeaTestSelPrio}.

\subsection{Target Systems and Data collection}
\label{system}


Our evaluation relies on two systems, namely \texttt{Libssh} and \texttt{Libsoup}, which further details  are presented in Table~\ref{table:target_systems}. These systems are also used in the HCS literature~\cite{ Medeiros2018, Oliveira2019}.  
\texttt{Libssh}\footnote{\url{https://www.libssh.org/}} is an open source cross-platform SSH library in C that implements the SSHv2 protocol on the client and server side. This library is designed to remotely run programs, transfer files, use a secure and transparent tunnel, manage public keys, to cite some of its features. \texttt{Libssh} is an HCS statically configurable with the C pre-processor directives. 
\texttt{Libsoup}\footnote{\url{https://wiki.gnome.org/Projects/libsoup}} is a client/server HTTP library developed for the GNOME environment. Utilizing GObjects and a simplified main loop, it seamlessly integrates with GNOME applications and provides a synchronous API designed for ease of use in command-line interface (CLI) tools. Its key features include asynchronous APIs, automatic connection reuse, TLS support, proxy functionality, compatibility with HTTP/1.1 and HTTP/2, and Client support for Digest, as well as  Server support for Digest. 

\begin{table*}[!htp]
\caption{Target Systems Information}
\label{table:target_systems}
\addtolength{\tabcolsep}{-2pt}
\centering
\small
\begin{tabular}{l|c|c|c|c|c|c}
\hline
\textbf{Systems} & \textbf{Fea-} & \textbf{Total} & \textbf{Commits} & \textbf{Commits} & \textbf{LOC} & \textbf{Contrib-} \\ 
 & \textbf{tures} & \textbf{Commits} & \textbf{Used} & \textbf{With Logs} &  & \textbf{utors} \\ 
\hline
\hline
\texttt{$Libssh$} & 144 & 5,958 & 4,388 & 303 & 85,817 & 110 \\ 
\hline
\texttt{$LibSoup$} & 29 & 3,896 & 3,420 & - & 64,265 & 247 \\ 
\hline
\end{tabular}
\end{table*}

We used PyDriller~\cite{Spadini2018} to mine the commit history of the HCS and  find the files changed in each commit. These files are the source of information to identify features changed (Step~2: Identify feature changed) and also to create the traceability between such files and test cases. Based on these trace links, we selected the test cases that are used by the changed-file-oriented approach. For that, the source code is scanned, looking for the test files. 

For the data collection, \texttt{Libssh} repository contains almost six thousand commits, from which we used 4,388. We discarded commits not associated with test cases and commits that are fork outside the repository. This set will be referred as the {\it whole set of commits}. To evaluate the quality of prioritized test cases, we need logs with failure and runtime information. To this end, we used a repository containing 303 commits from a related work~\cite{lima2020learning}, referred as {\it set of commits with logs}. The granularity of the logs is file-level, which gives us information about which test files reveal failures, called herein as failed files. 
For \texttt{Libsoup}, we could not obtain information about the  logs, which did not allow us to know which test cases failed in each CI cycle. Thus, for \texttt{Libsoup} we used 3,420 commits in our experiments, out of almost four thousand commits, equivalent to the {\it whole set of commits}. We also discard commits that do not have associated test cases and commits that are fork outside the repository. 
Due to the impossibility of obtaining CI logs for \texttt{Libsoup}, only \texttt{Libssh} was used in the analyses to answer RQ2 to RQ4, which require  the execution time of each test case, as well as whether it failed, information included in the commits with logs.

\subsection{Evaluation Measures}
\label{sec:measures}

To compute the average execution time of each test case, an approximation is necessary, since the evaluated approaches use function-level granularity (i.e., a test case corresponds to a function) and the system logs use file-level granularity. Thus, for each CI cycle, there is an execution time per test file for each job of that cycle. To perform this approximation, we first  mined the number of functions each test file has and the average time this file takes to execute. We then divided the average time by the number of functions to compute the average time each test case (i.e., function) takes to run. For example, if a file takes an average of 100 seconds to execute in a commit and has 10 functions, each function takes an average of 10 seconds to execute. Then, we compared how many functions were selected by each approach for each test file. For example, if our approach selects 5 functions from these files, it would take 50 seconds on average to execute, which represents a 50\% reduction in time compared to the execution time of the complete file.

To validate the quality of the approaches, we use the criterion based on detected failures. To this end, we mined the logs to identify test files failed in the failed commits. For instance,  the commit \texttt{10728f85}\footnote{See Log Line 1130 at \url{https://gitlab.com/libssh/libssh-mirror/-/jobs/78303050}} has three failed test files: \texttt{torture\_packet}, \texttt{torture\_algorithms}, and \texttt{pkd\_hello}, thus, three failures are counted. We perform an analysis per-commit to verify whether the test cases selected by both approaches cover the failed test files. For example, considering the previous commit, if the approaches select test cases from the three files, the three failures are considered as detected. However, if only the \texttt{pdk\_hello} file is affected, then only one failure is  detected.

To answer RQ4, we adopted  an early failure detection indicator. This indicator quantifies the average \textit{test budget} required, that is, the percentage of test cases to be executed to find a failure in a commit. To identify each test case that fails in the commit, we rely on the system log. We compute the average budget required per commit because for each commit pipeline, there can be multiple jobs. Consequently, test cases run in parallel across these jobs, allowing different test cases to fail in the same commit.

Equation~\ref{eq:budget} presents how we compute the early failure detection indicator. The average budget required to identify the failure is determined by $\gamma$, providing the average percentage needed to locate the test cases that fail in the commit. This equation is based on the sum of the position ($P_i$) of the test cases in the prioritized list, where $nt$ is the total number of test cases for the commit, and $nf$ is the total number of test cases that fail in the commit. $\gamma$  represents the average budget per commit to identify the fault, considering all jobs. 

\begin{equation}\label{eq:budget}
{\gamma} = \frac{\sum_{i=1}^{nf} P_i \text{nt}}{{100*nf}}
\end{equation} 

For example, in the commit \texttt{17b518a}, there are 27 jobs, and  failures occurred in two of these jobs. In the job \texttt{visualstudio/x86\_64},\footnote{\url{https://gitlab.com/libssh/libssh-mirror/-/jobs/432862254}} the test case \texttt{torture\_rekey} failed. Meanwhile, in the job \texttt{ubuntu/openssl\_1.1.x/x86\_64},\footnote{\url{https://gitlab.com/libssh/libssh-mirror/-/jobs/432862274}} the failure occurred in the test case \texttt{torture\_misc}. To locate the \texttt{torture\_misc} test file in this commit, it is necessary to execute 34 test cases from the total set of 638, which is approximately 5.3\%. For the \texttt{torture\_rekey} file, approximately 2.1\% (13) of test cases are required. Therefore, to assess the commit as a whole, we compute the mean percentage, resulting that in average, 3.7\% of test cases need to be executed to find both failures.

\subsection{Platform for the Experiments}

All experiments were conducted on a server equipped with an Intel(R) Xeon(R) Gold 6230N CPU @ 2.30GHz 48-Cores, 252GB RAM, running on Linux Ubuntu 18.04.6 LTS.


\section{Results and Analysis}
\label{result}

This section presents the results and discussions to answer the RQs of our study. The dataset and results are available online~\cite{Mendonça_Vergilio_Assunção_2024}, as a supplementary material.

\subsection{RQ1: Performance of \texttt{FeaTestSel} and changed-file-oriented approaches}
\label{answerRQ2}

To answer RQ1, we compare the \texttt{FeaTestSel} and changed-file-oriented approaches against retest-all, regarding the number of selected test cases. 
In this comparison, we analyze the percentage of reduction in the number of test cases and execution time. As both \texttt{FeaTestSel} and changed-file-oriented approaches consider function granularity, we count the number of test functions selected, as mentioned in Section~\ref{sec:measures}.  We consider the number of test cases to be executed by the retest-all  strategy is given by the number of functions existing within all the files in the logs. Then, each function is considered a test case. 

Table~\ref{tab:reduction} presents, for both approaches, the percentage of reduction in the number of test cases in comparison to the retest-all technique, and time taken to perform the selection (in seconds). For \texttt{Libssh}, we consider both sets of commits (with or without logs).  The minimum reduction percentage  for both systems is 0, which occurred in the initial commits where the number of test cases is small, and in several commits, it is not possible to reduce the set because  usually 100\% of the test cases are selected.  We observe that the percentage of test reduction reached by  \texttt{FeaTestSel} vary according to the system and scenario considered. For  \texttt{Libsoup} the percentage is lower. A possible explanation for this is on the number of features. \texttt{Libsoup} is smaller and has a reduced number of features, only 29, against 144 of \texttt{Libssh} (see Table~\ref{table:target_systems}).  In this case, many test cases are associated with the feature \texttt{BASE} and are selected in the majority of commits. 

The reduced number of features also implies a shorter time to perform the selection (11.47 seconds), making the feature location step less time-consuming. But this difference does not grow proportionally.  \texttt{Libssh} has almost five times the number of features than \texttt{LibSoup}, but it requires only the double of time to execute (20.82 seconds).

Another point to be evaluated is the percentage of reduction considering the set of commits evaluated. We observe for \texttt{Libssh} that the percentage of reduction is greater when the set of commits with log is considered. In such scenario, there is no case where the percentage is zero. The 303 commits with log does not belong to the set of initial commits. These commits are in the middle of the evolution process of the system. When the system evolves and becomes more complex (in the last commits) there are a greater number of test cases, then the results of using a TCS approach are more significant.  We also observe this difference regarding the set of commits considered for the changed-file-oriented approach.

\begin{table*}[!htp]
\caption{Percentage of Reduction in the Number of Test Cases in Comparison to Retest-All and Time to Perform TCS -  \texttt{FeaTestSel} and Changed-file-oriented Approach}
\label{tab:reduction}
\addtolength{\tabcolsep}{-4pt}
\centering
\small
\begin{tabular}{l||r|r|r|r|r|r||r|r|r|r|r|r} 
\hline
\multirow{3}{*}{\textbf{System}} & \multicolumn{6}{c||}{\textbf{{\tt FeaTestSel} }} & \multicolumn{6}{c}{\textbf {Changed-File Oriented Appr.}} \\ 
\cline{2-13}
& \multicolumn{3}{c|}{\textbf{\% Test Reduction}} & \multicolumn{3}{c||}{\textbf{Time (seconds)}} & \multicolumn{3}{c|}{\textbf{\% Test Reduction}} & \multicolumn{3}{c}{\textbf{Time (seconds)}} \\ 
\cline{2-13}
& \textbf{min} & \textbf{max} & \textbf{avg} & \textbf{min} & \textbf{max} & \textbf{avg} & \textbf{min} & \textbf{max} & \textbf{avg} & \textbf{min} & \textbf{max} & \textbf{avg}  \\ 
\hline
\hline
\texttt{$Libssh^*$}  & 0 & 45.29 & 24.22 & 9.68 &35.72 & 20.82 & 27.5 & 100 & 94.23 & 2.01 & 21.04 & 8.73\\ \hline 
\texttt{$Libssh^\$$} & 37.74 & 45.29 & 41.98 & 18.61 & 35.72 & 25.97 & 51.11 & 100 & 97.27 & 6.77 & 18.41 & 11.98  \\ \hline  
\texttt{$LibSoup$}   & 0  & 39.06 & 18.62 & 4.15  & 37.68 & 11.47 & 0 & 100 & 91.68 & 0.03 & 8.40 &  4.83 \\
\hline  
\end{tabular}

{\small * whole set of commits; \$ set of commits with log}
\end{table*}

The average percentage reduction in test cases for our approach, considering all three configurations (i.e., $Libssh^*$, $Libssh^\$$, and $LibSoup$), is equal to 21.41\%, whereas for the changed-file-oriented approach, it is 92.95\%. Notice here that at this point we are not considering the quality of test cases selected. The average runtime for our approach is 16.14 seconds, whereas for the changed-file-oriented approach, it is 6.78 seconds. As expected, our approach takes longer to execute due to feature to test case traceability step.

\begin{figure}[!t]
	\begin{minipage}[c]{1\linewidth}
		\centering
		\subfigure[\texttt{$Libssh^*$}]{
			\label{fig:numberTestW}
			\includegraphics[width=.8\textwidth]{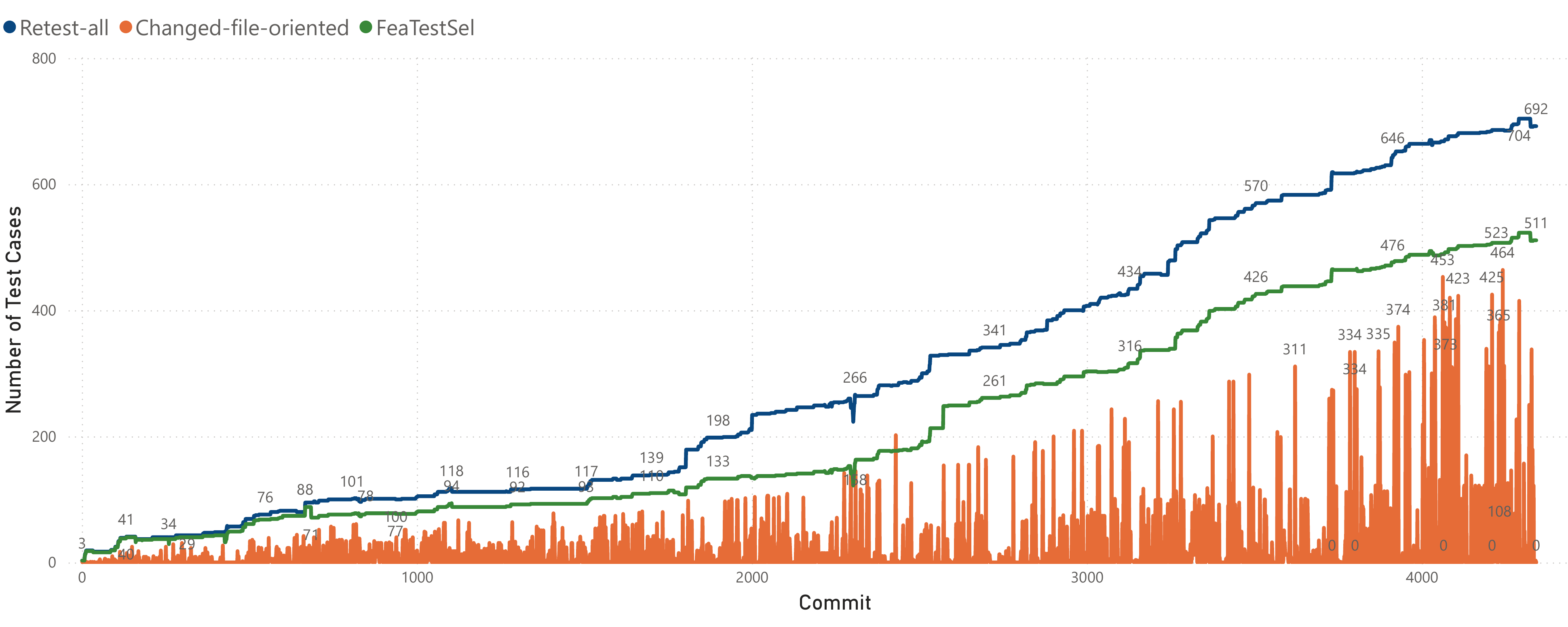}
		}
	\end{minipage}\\%
	\begin{minipage}[c]{1\linewidth}
		\centering
		\subfigure[\texttt{LibSoup}]{
			\label{fig:numberTestSoup}
			\includegraphics[width=.8\textwidth]{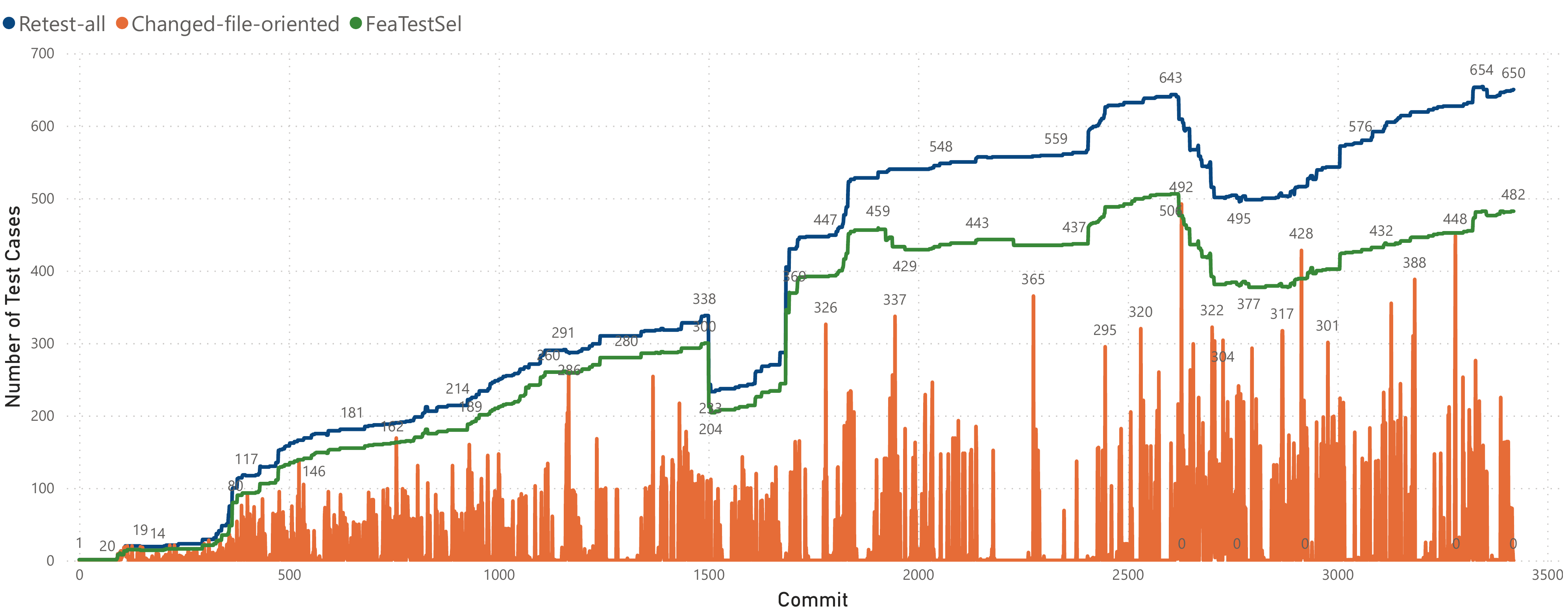}
		}
	\end{minipage}%
	\caption{Number of test cases selected by the approaches for the whole set of commits}
	\label{fig:RQ1-nrtest}
\end{figure}

To analyze the behavior of the approaches over the commits, we present the absolute number of test cases selected for each commit in Figure~\ref{fig:RQ1-nrtest}. 
The blue line represents the number of test cases available for the commit (retest-all technique); the green line represents the test cases selected by our approach; and the orange line represents the number of test cases selected by the changed-file-oriented approach. We can observe that for both systems, the changed-file-oriented approach selects fewer test cases than \texttt{FeaTestSel} in the great majority of commits. However, in some instances, it does not select any test case, as indicated by the orange line in the figure.

\begin{figure}[!htp]
  \centering
  \includegraphics[width=1\linewidth]{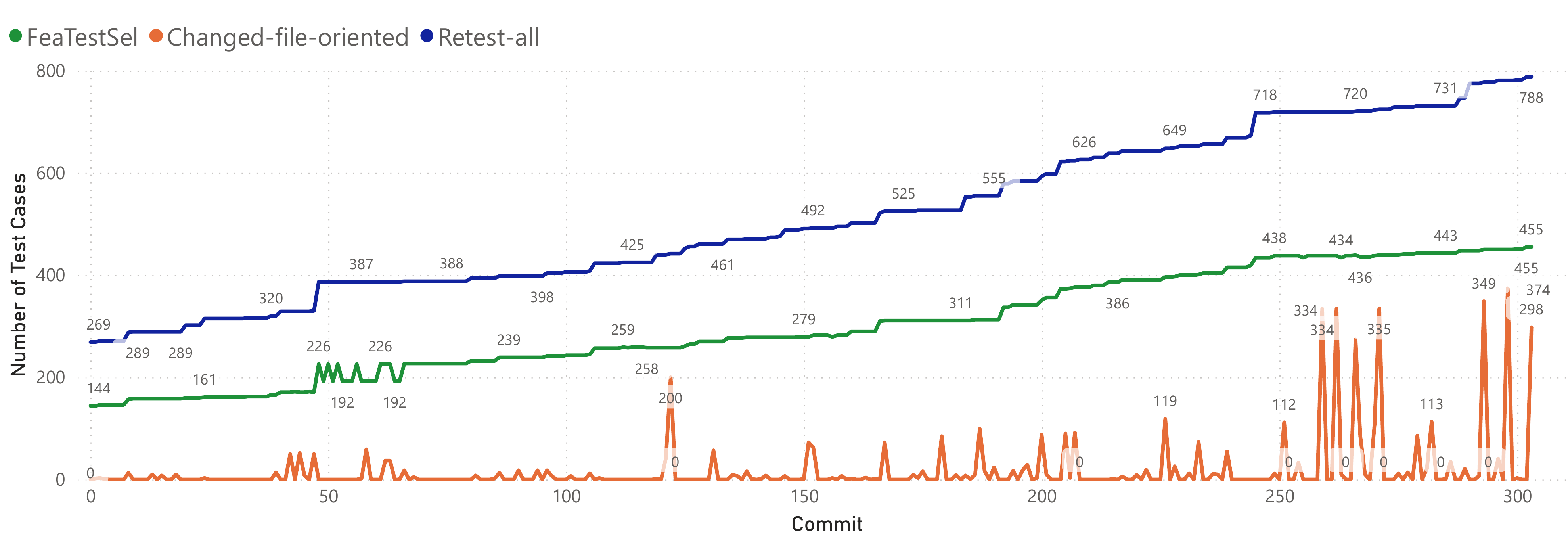} 
  \caption{Number of test cases selected by the approaches for the set of commits with logs - \texttt{$Libssh^\$$}}
  \label{fig:numberTestCL}
\end{figure}

To better compare both approaches and consider the execution time of each test case, we use the system \texttt{Libssh} considering the 303 commits with logs.  We first generate Figure~\ref{fig:numberTestCL}. 
that  shows the number of test cases selected by each approach in each commit. 
The blue line shows the number of test cases available for the commit (retest-all); the green, the test cases selected by our approach; and the orange, the number of test cases selected by the changed-file-oriented approach. For this set, we observe that the file-oriented approach always selects fewer test cases than \texttt{FeaTestSel}. However, in several commits, it does not select any test cases, as indicated by the orange line in the figure.
This happens in 199 out of 303 commits ($\approx$65.67\%). In other commits a few test cases are selected, in fact the number of test cases selected depends on the performed changes. For example, in commit \texttt{12284b75},\footnote{\url{https://gitlab.com/libssh/libssh-mirror/-/commit/12284b75}} which involves changes in six files, only eight test cases were selected from a single test file, out of a total of 329 test cases available. But there are cases where a significant number of changes are made, and many test cases are selected, as it happens in commit \texttt{17b518a6}.\footnote{\url{https://gitlab.com/libssh/libssh-mirror/-/commit/17b518a6}} In this commit seven files are changed (245 additions and 13 deletions), and this approach selected 334 out of 719 test cases. This does not happen for our approach, which  selects test cases in all commits.

\begin{figure}[!htp]
  \centering
  \includegraphics[width=.8\linewidth]{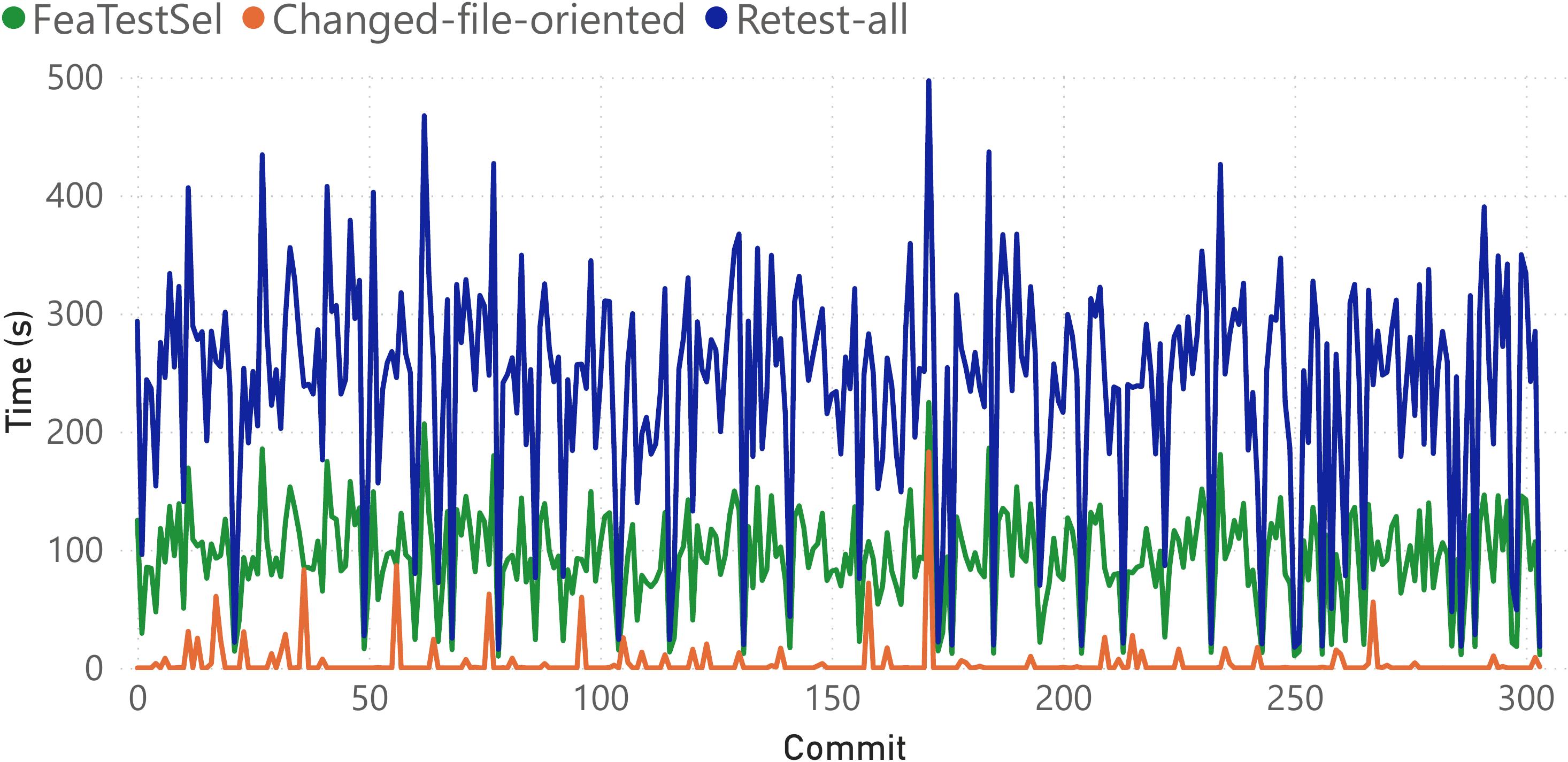}
  \caption{Average time to execute the selected test cases per commit and approach for the set of commits with logs -  \texttt{$Libssh^\$$}}
  \label{fig:duration}
\end{figure}

When we apply a TCS approach, we may be looking for a way to reduce the time spent executing the test cases.  Figure~\ref{fig:duration} shows the average time spent per commit with the execution of the  test cases selected by each approach. We observe that the orange line is always lower, showing that the test set selected by the changed-file-oriented approach always takes less time to execute. The average time to execute all test cases (retest-all) is 239.22 seconds. The time to execute the test cases selected by \texttt{FeaTestSel} is 92.78 seconds,  and the time to execute the ones selected by changed-file-oriented approach is  4.15 seconds. We can observe that this time reduction is quite drastic, what may affect the quality of test cases for failure detection, subject investigated in RQ2.

\answer{1}{The changed-file-oriented approach leads to a expressive reduction in the number of test cases compared to our approach (average of 92\%), as well as in the time to execute the selected test cases (4.15 seconds). This reduction depends on the changes in the commit, and the evolution process. The reduction percentage achieved by our approach (average of 21.41\%) varies based on the system's size, number of tests available in the commit (evolution process), and number of features. In the set of commits with logs, this percentage reaches $\approx$42\%, and the mean time to execute the selected test cases is 92.78 seconds.}


\subsection{RQ2: Quality of selected test cases for failures detection}
\label{answerRQ3}

This RQ evaluates the quality of the test cases selected by \texttt{FeaTestSel} against the other approaches. The goal is to analyze if the reduction in the test cases impacts the number of failures detected. To this end, we use the set of \texttt{Libssh} commits with logs.

\begin{figure}[!htp]
  \centering
  \includegraphics[width=.8\linewidth]{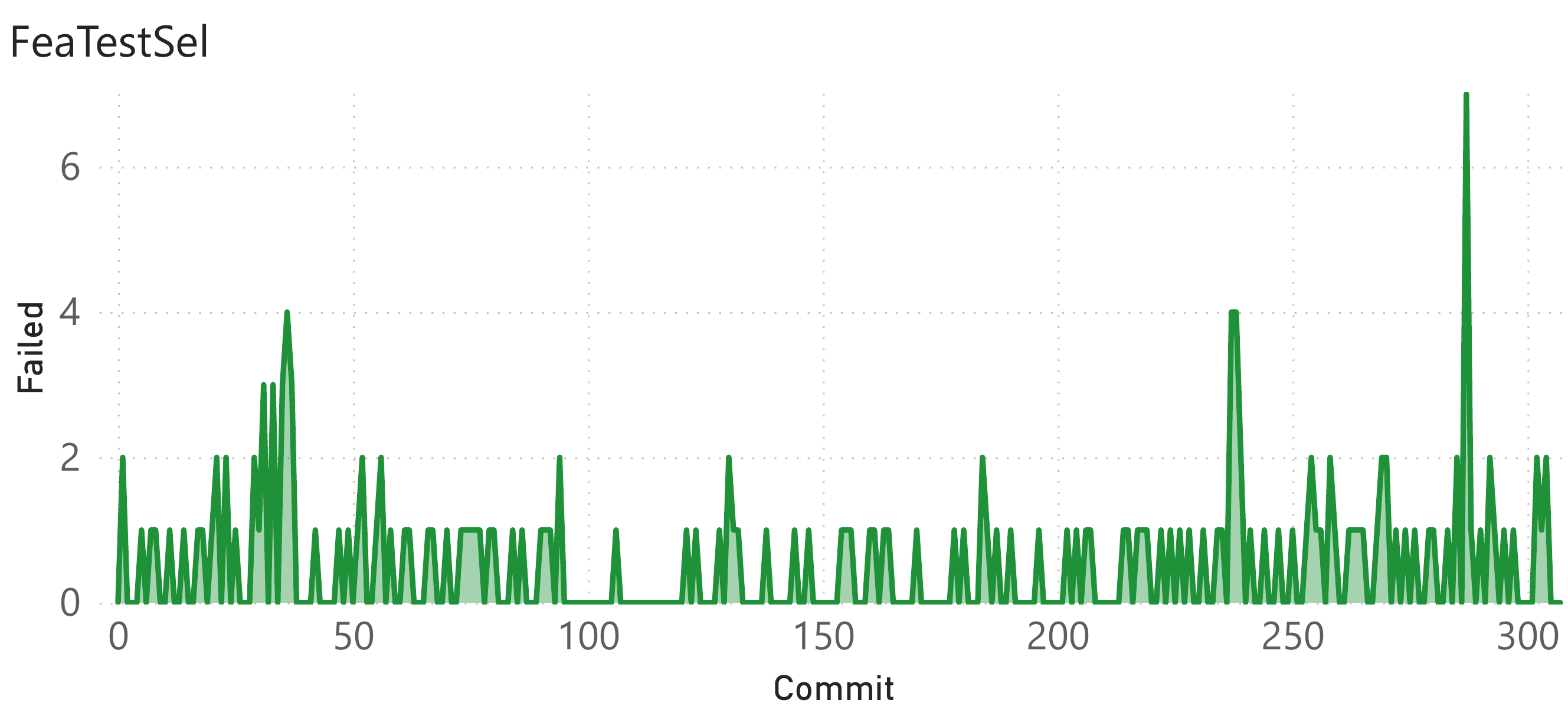}
  \includegraphics[width=.8\linewidth]{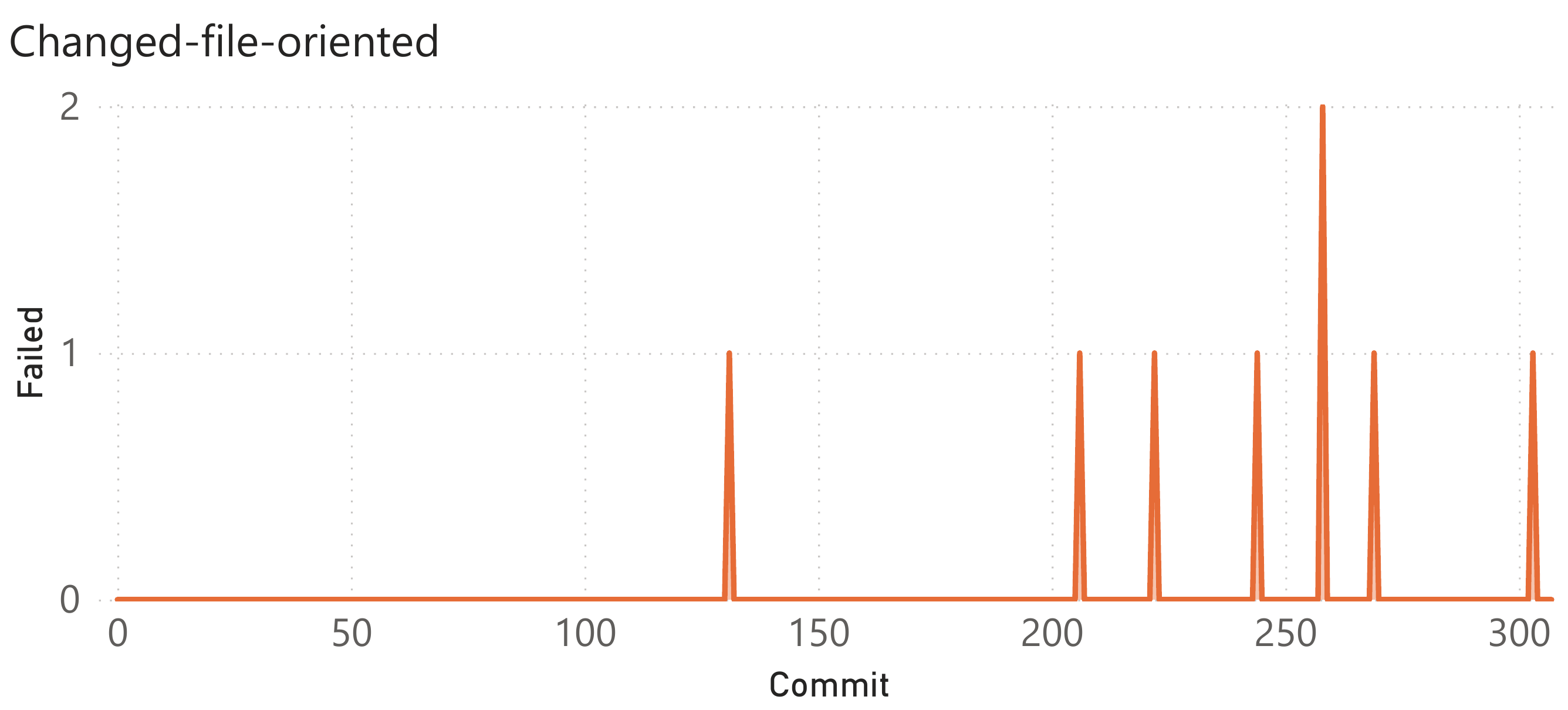}
  \includegraphics[width=.8\linewidth]{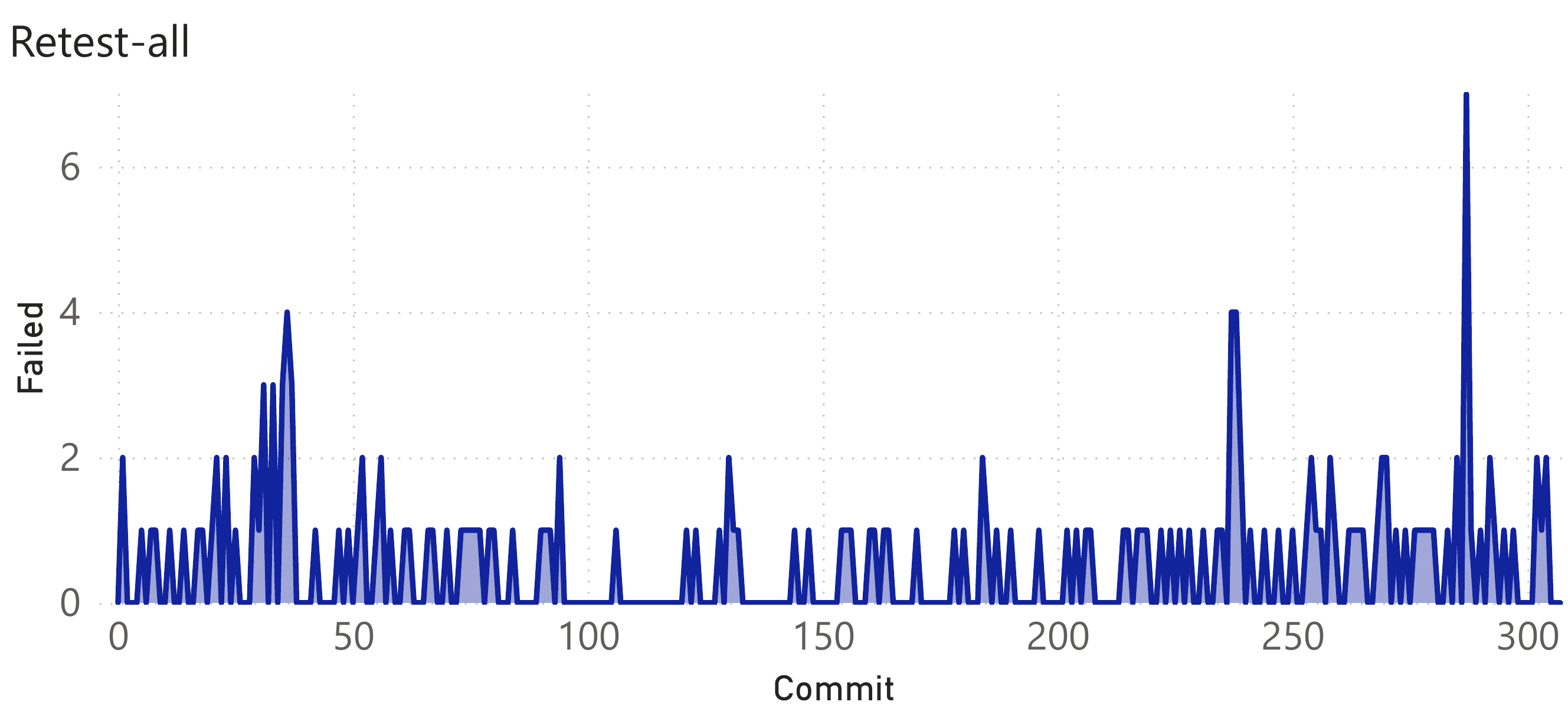}
  \caption{Number of failed files per commit and approach for the set of commits with logs}
  \label{fig:Fail}
\end{figure}

Figure~\ref{fig:Fail} presents the number of failures detected with the test cases selected. We can observe that the curves corresponding to our approach  (green line, at the top of the figure) and retest-all (blue line, at the bottom) are exactly the same, which does not happen for the changed-file-oriented approach (orange line, in the middle). This means that the changed-file-oriented approach has a poor performance for selecting test cases that detect failures. This happens because the changed-file-oriented approach selects only a small set of test cases related to the files changed, as showed in the previous section. Despite leading to an expressive reduction of the number of test cases selected and execution time, these test cases do not maintain the quality of the testing activity. 

\begin{figure}[tp]
  \centering
  \includegraphics[width=.6\linewidth]{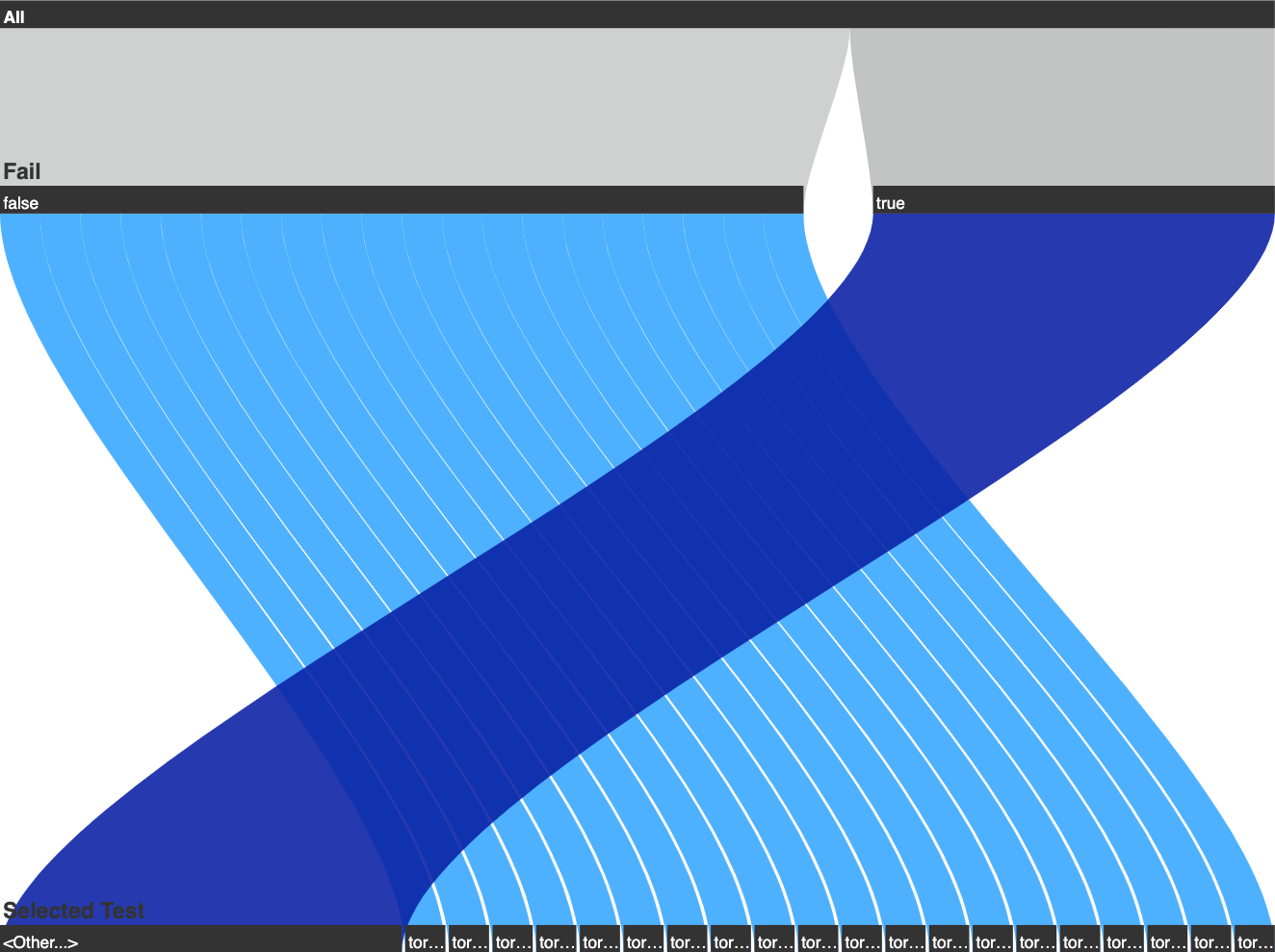}
  \caption{Failed and selected test cases for commit \texttt{d7477dc}}
  \label{fig:FailExample}
\end{figure}

To complement the analysis, we created a dynamic chart\footnote{Available at \href{https://app.powerbi.com/view?r=eyJrIjoiYjk3MGUzNTgtOWE5MS00ZWNhLWI1MGMtZTYzMTcwZDVlODZlIiwidCI6ImRhZGFhOGQzLTIxYWEtNGRjNS05ODBlLTFiZjI0ZWY5Yzc0OCJ9&pageName=ReportSection}{https://app.powerbi.com/view?...}} 
considering the 303 commits with logs. Figure \ref{fig:FailExample} shows a clipping of the graph for the commit~\texttt{d7477dc}.\footnote{\href{https://gitlab.com/libssh/libssh-mirror/-/commit/d7477dc}{https://gitlab.com/libssh/commit/d7477dc}}   The graph is separated into three main lines: the first line represents all test cases; on the left part of the second line the failed test cases (false), and on the right of this line the test cases that passed (true); and the third line represents the test cases selected by our approach.  In this commit, 162 test cases were selected by our approach from a total of 316 available.  We can see for this commit that all the failed test cases were selected by our approach, so we are keeping the quality, considering the failure criterion. We can see similar behavior for the other commits in the dynamic chart made available.

\answer{2}{Based on the failure criterion and considering the retest-all technique baseline, our approach manages to always select (i.e., 100\% of the times) the test files that reveal failures, maintaining the quality of the test case set. This does not happen for the changed-file-oriented approach.}

\subsection{RQ3:  \texttt{FeaTestSelPrio} applicability}
\label{answerRQ1}


This RQ investigates the applicability of our approach, considering the time available in the CI cycles. To answer RQ3, we need the time each test case takes to execute, thus, the results rely only on \texttt{Libssh} and its 303 commits with logs. For this analysis, we evaluate \texttt{FeaTestSel} applicability considering  the time it takes to execute, summed to the time to execute the selected test cases. 

As mentioned in the last section, the average time to execute all the test cases is 239.22 seconds ($\approx$4 minutes). The test cases selected by \texttt{FeaTestSel} take on average 92.78 seconds to run. By adding to this time the average time our approach takes to perform the selection (25.97 seconds), the runtime is 118.75 seconds ($\approx$2 minutes). This represents a reduction of $\approx$50\% in the total runtime compared to the retest-all technique. We also observe that the interval between the CI cycles is on average 1142.52 minutes (standard deviation equals to 459.28 minutes), what shows the applicability of our approach.

\answer{3}{ When using the \texttt{Libssh} commits with logs, the average time to perform the selection is 25.97 seconds. Summing this time to the time spent to execute the selected test cases, the total is 118.75 seconds. This represents a reduction of $\approx$50\% compared to retest-all. This time is compatible with the time between CI cycles, what shows the applicability of \texttt{FeaTestSel}.}

\vspace{0.1cm}

\subsection{RQ4: Evaluating {\tt FeaTestSelPrio}}
\label{answerRQ4}

To answer RQ4, we evaluate \texttt{FeaTestSelPrio} in comparison to \texttt{FeaTestSel} considering early failure detection. For this end, we use the set of commits with log of \texttt{Libssh} and the budget required to detect a failure  (see Section~\ref{sec:measures}). As mentioned before, \texttt{FeaTestSel} does not produce any order, then we execute the selected test cases according to the order they appear in the output of our implementation.  

In Table \ref{tab:budgetmean}, we can observe that the average budget value of \texttt{FeaTestSelPrio} is almost half of the budget value of \texttt{FeaTestSel}. The difference in the minimum values is very small, but it is significant in the maximum ones. Also, \texttt{FeaTestSelPrio}  has a lower standard deviation (SD) than  \texttt{FeaTestSel}, which represents that the set of ordered test cases improves the dispersion of the budgets necessary to the failure.

\begin{table}[!htp]
\caption{Comparison between {\tt FeaTestSelPrio} and {\tt FeaTestSel} regarding the number of test cases required (budget) to detect known failures}
\label{tab:budgetmean}
\centering
\begin{tabular}{l|l|l|l|l}
\hline
\multicolumn{1}{c|}{\textbf{Strategies}} & \multicolumn{1}{c|}{\textbf{Avg}} & \multicolumn{1}{c|}{\textbf{Max}} & \multicolumn{1}{c|}{\textbf{Min}} &  \multicolumn{1}{c}{\textbf{SD}} 
\\ 
\hline
\hline
{\tt FeaTestSel} & 44.99\% & 94.31\%  & 0.67\%  & 34.68 \\ 
\hline
{\tt FeaTestSelPrio} & 23.4\% & 71.82\% & 0.56\% & 16.37 \\ 
\hline
\end{tabular}
\end{table}

\begin{figure}[!htp]
  \centering
  \includegraphics[width=.95\linewidth]{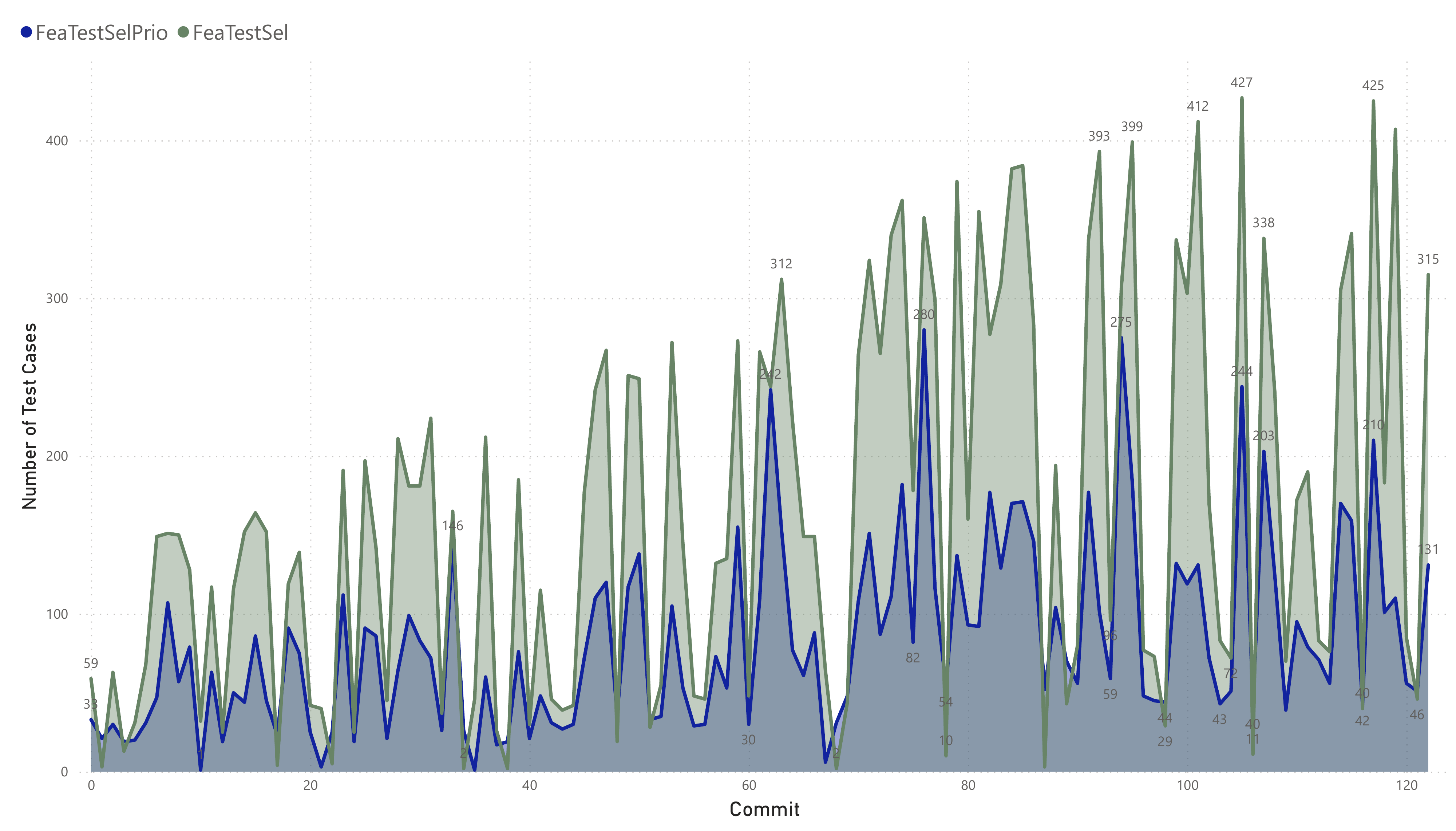}
  \caption{Comparing approaches according to number of test cases executed to find a failure in the failed commits.}
  \label{fig:testnumber}
\end{figure}

The results indicate that the use of \texttt{FeaTestSelPrio} leads to an early detection of failures with lower budgets on average. To corroborate our analysis, we compared the \texttt{FeaTestSel} and \texttt{FeaTestSelPrio} approaches according to the number of tests executed until a failure is produced,  and generated Figure~\ref{fig:testnumber}, which displays a graph for the 122 failed commits from the total of 303 commits. The green line presents the number of test cases executed by \texttt{FeaTestSel} approach and the blue line by \texttt{FeaTestSelPrio}, which is mostly below the green line for all commits.  It is possible to identify the failure, that is, to reduce budget, with a smaller number of test cases using \texttt{FeaTestSelPrio} in 86\% of the commits.

The mean time to perform the prioritization step of \texttt{FeaTestSelPrio} is 0.42 seconds (minimum of 0.37 seconds, maximum of 0.74 seconds, and standard deviation equal to 0.03 seconds). This time summed to the time \texttt{FeaTestSel} takes to execute is equal to 26.40  seconds. Finally, we have to sum the time to actually execute the selected test cases, resulting in a total of 98.78 seconds, which is the worst case in the prioritization rank. This total time of 98.78 seconds is acceptable for practical application of our approach, considering the duration of the CI cycles.

\answer{4}{ The results of this RQ highlight the importance of the additional prioritization step in nearly every commit. The approaches \texttt{FeaTestSel} and \texttt{FeaTestSelPrio} identify a fault  with a mean budget of, respectively, $\approx$44\% and  $\approx$23\%. The incorporation of a prioritization step results in a reduction of $\approx$50\% in the average budget required. Furthermore, \texttt{FeaTestSelPrio} allows fault detection earlier than \texttt{FeaTestSel} in $\approx$86\% of the faulty commits, requiring an additional mean time of 0.42 seconds.}

\section{Discussion}
\label{discussion}

In this section, we discuss some implications as well as the threats to validity and limitations of our approach and evaluation. 


\subsection{Implications for the research and practice}

\texttt{FeaTestSelPrio} is lightweight. Differently from other TCS approaches, our approach does not require neither a failure-history nor the application of search-based/learning techniques, what leads to a reduced time to perform the selection and prioritization.  It is noteworthy to observe that the approach produces an entire static analysis of the system for each commit, delivering results that can be analyzed qualitatively and/or quantitatively by developers for possible improvements in the system. For example, developers can use the test case traceability to feature to identify features that need more test.  

By analyzing the whole set of commits, we observe that the time the approach takes to execute is dependent on the size of the system and number of test cases, as well as the percentage in the test reduction. This implies that the approach is more beneficial as the system evolves and a greater number of test cases are in the repository. However, this relationship should be better explored in future works. 

In comparison with a changed-file-oriented approach, \texttt{FeaTestSel} presents some advantages, obtaining a good balance in the test set reduction, making sure that some important test cases are selected in the regression testing, maintaining quality regarding failure detection. The reduction provided by the changed-file-oriented approach is very dependent on the number of changes in the commit. When a CI environment is adopted in the development, it is a very common practice to perform  small changes and push them to the repository. This may be the reason why the changed-file-oriented approach usually selects a small test set, or even no test cases at all. This makes our approach more suitable in this scenario. 

We observe that our approach takes only few seconds to execute, and is suitable for the CI context. The time of the test cases selected take to execute summed to the time taken to perform the selection, or the selection and prioritization, is lower than the CI cycles, what shows  the applicability of the approach. 
\texttt{FeaTestSel} contributes to select a reduced number of test cases, and consequently reduces the test execution time.  We observe that it fits in a budget of 50\% of time that would be spent by executing all the available test cases for the commit without loosing quality regarding possible failures produced.  However, as RQ4 pointed out the use of \texttt{FeaTestSelPrio} is very important in the presence of a time budget as it happens in the CI scenarios, allowing an early fault-detection and a rapid feedback to the tester. In this work, we explored only a prioritization strategy based on the feature coverage. But other strategies could be explored in future work. Such strategies can explore other test characteristics such as test complexity, user, preferences, severity of the failures they reveal, feature-change history, and so on.

A limitation for the use of our approach is that our implementation requires as input HCSs developed using pre-processor directives (i.e., an annotative approach). Despite this being a limitation, the use of pre-processor directives is a very common practice, adopted by most HCSs~\cite{Medeiros2018}. However, using other variability mechanisms such as component technologies~\cite{szyperski2002component},
frameworks~\cite{johnson1988designing}, or feature-oriented programming with some form of
feature modules~\cite{prehofer1997feature} (i.e., a compositional approach) should be the focus of future work~\cite{moreira2022open}.

\subsection{Threats to Validity}
\label{TheVali}

In this section, we discuss some treats to the validity of our study according to classification by Wohli et al.~\cite{Wohlin2000}.

\noindent \textit{Internal Validity:} We faced a problem when dealing with code scanning tools, which often use global variable names in their traceability. This lead to traceability false positives between global and local variables with the same name. To resolve this issue, we dropped all global variables during the {\it Map feature to Test Cases} step. To further improve this, we intend to completely eliminate the consideration of global variables in traceability during the code scan task. Furthermore, in the \textit{Prioritize test cases} step, we consider the count of the number of features covered by the test case. However, in cases of a tie between test cases, the choice is currently made randomly. We plan to incorporate tie-breaking strategies, such as considering feature changes and the feature change history, to enhance this process.  Moreover, it should be noted that the \texttt{Libssh} system logs use a file granularity, which led us to  calculate an approximated value for the execution time of each test function in the test files. This may affect the accuracy of the test execution time calculation, but we still ensure that the total test suite is reduced while maintaining failure-based quality.

\vspace{0.12cm}
\noindent {\it Construct validity:} A threat in this category is the approach used in the comparison. As we did not find approaches or tools available for the C/C++ language, we used a simple but well-known changed-file selection approach that we developed internally as a basis for comparison. Possible errors in the implementations were minimizing by a manual validation of the results produced. Additionally, for the prioritization step, we performed a comparison with \texttt{FeatTestSel}, which produces a list of selected test cases in alphanumeric order. This could potentially impact the results. Furthermore, experiments should be conducted to evaluate alternative prioritization strategies.

\vspace{0.12cm}
\noindent \textit{External validity:}  We used only two systems in our evaluation, and only \texttt{Libssh} for answers the RQs that needed error log of the test cases. This is a threat, and our results may not be generalized. For future work, other industrial-grade systems should be included in the validation. Our implementation currently works for different annotated systems. But we have not found systems with logs. To minimize this threat, future work can execute other subject systems to report the test execution results, improving generalization.

\vspace{0.12cm}
\noindent \textit{Conclusion validity}: a possible threat in the analysis conducted in our study is the indicators used, which may impact the results. To minimize this, we adopted indicators that are relevant for evaluating test case selection and prioritization techniques, and are related to reduction in the number of test cases and early-fault detection. 

\vspace{0.12cm}
\noindent \textit{Reliability validity} is concerned with reproducibility. We believe our study is replicable by following the steps outlined in Section \ref{methodology}, and we have made all raw results and logs available in our repository.



\section{Concluding Remarks}
\label{conclu}

This work introduces  \texttt{FeaTestSelPrio}, a feature-oriented test case selection and prioritization approach to be applied during the evolution of HCSs. Unlike other selection approaches that mainly focus on test cases directly related to changed files without considering features, our approach takes into account interactions with other parts of the features being modified in a given commit.  The prioritization step uses the feature covered to rank test cases. Although we use system logs to validate our approach, we do not need access to any type of failure log or failure history for execution. Differently from several exiting prioritization approaches, the failure-history or dynamic analysis is not required.  \texttt{FeaTestSelPrio} uses as input basically only the source code of the HCS and a configuration file as input.
The approach produces several intermediate reports as output, which can be used by software engineers to make decisions for maintenance and/or upgrades for their HCSs. These reports allow  an overview  of various aspects of the systems, which are not possible to be analyzed only using the source code.

In the analysis of the set of commits with logs, our feature-oriented selection reaches an average reduction in the number of test cases of $\approx$42\%. The time it takes to execute summed to the time to execute selected test cases fits well in a budget of 50\%. 
These percentages depend on the system size, number of features and test cases, being more significant in the last commits, when the system evolved and has a greater number of test cases. This reduction does not imply loosing important test cases because the approach manages to select 100\% of the failed test files, maintaining the test quality.
\texttt{FeaTestSelPrio} is applicable considering CI cycles.  When the set of commits with logs from \texttt{Libssh} is considered, the average time to perform the selection and prioritization is 26.40 seconds. This time, when summed to the time spent in the worst case to execute all the selected test cases, is 119.75 seconds. To execute the prioritization step of \texttt{FeaTestSelPrio} an average time of only 0.42 seconds is required. 

Overall, our approach selects a greater number of test cases and takes longer to execute than  a changed-file-oriented approach, but we observe a greater quality regarding the failures detected. We also observe the importance of the prioritization step that allows a reduction in the average budget in 86\% of the failed commits. 

Future work includes to obtain test logs of other systems and conduct other experiments. We intend to better evaluate the relationship between the system size and the reduction in the number of test cases, as well as execution time. Furthermore, we intend to improve the selection by adding some criteria, such as changes in files along with changes and features. Additionally, we should consider using Artificial Intelligence algorithms for a potential comparison and/or improvement of the results. Another research direction  involves  to propose and evaluate other prioritization strategies, which are also feature-oriented, but that consider detected and severity failures.


\section*{Acknowledgment}
This research was funded by CNPq (Grant 310034/2022-1)), FAPERJ PDR-10 program (Grant 202073/2020), and CAPES (Grant 88887.464736/2019-00).


\small
\bibliographystyle{elsarticle-num} 
\bibliography{references}
\end{document}